\documentclass[journal]{IEEEtran}

\usepackage{savesym}
\usepackage{tabularx}
\savesymbol{swdefault}
\usepackage{allrunes}
\restoresymbol{RUNES}{swdefault}
\usepackage[]{algpseudocode}
\usepackage{algorithm}
\usepackage{hyperref}
\usepackage{multirow}
\usepackage{booktabs}
\usepackage{subfig}
\usepackage[export]{adjustbox}
\usepackage[most]{tcolorbox}
\usepackage{authblk}
\usepackage{adjustbox}
\usepackage{amsmath}
\usepackage{amssymb}
\usepackage{breqn}

\begin{document}

\title{The Concept of Semantic Value in Social Network Analysis: an Application to Comparative Mythology}

%
%
%
%
%
%
\author{Javier Fumanal-Idocin,~\IEEEmembership{Member,~IEEE}, Oscar Cord\'on  ~\IEEEmembership{Fellow,~IEEE},  Graçaliz Dimuro, Mar\'ia Min\'arov\'a, Humberto Bustince,~\IEEEmembership{Senior,~IEEE}
	
	\thanks{Javier Fumanal-Idocin, Graçaliz Pereira Dimuro and Humberto Bustince are with the Departamento de Estadistica, Informatica y Matematicas, Universidad Publica de Navarra, Campus de Arrosadia, 31006, Pamplona, Spain.
		emails: javier.fumanal@unavarra.es, gracalizdimuro@furg.br, bustince@unavarra.es}
	\thanks{Oscar Cordón is with the Dept. of Computer Science and Artificial Intelligence, Andalusian Research Institute ``Data Science and Computational Intelligence'' (DaSCI), University of Granada, 18071 Granada, Spain.}
	\thanks{Humberto Bustince is with the Institute of Smart Cities, Universidad Publica de Navarra, Campus de Arrosadia, 31006, Pamplona, Spain.}
	\thanks{Humberto Bustince is with the Laboratory Navarrabiomed, Hospital Complex of Navarre (CHN), Universidad Publica de Navarra, IdiSNA, Irunlarrea 3. 31008, Pamplona, Spain.}

}

\maketitle

\begin{abstract}
Human sciences have traditionally relied on human reasoning and intelligence to infer knowledge from a wide range of sources, such as oral and written narrations, reports, and traditions. Here we develop an extension of classical social network analysis approaches to incorporate the concept of meaning in each actor, as a mean to quantify and infer further knowledge from the original source of the network. This extension is based on a new affinity function, the semantic affinity, that establishes fuzzy-like relationships between the different actors in the network, using combinations of affinity functions. We also propose a new heuristic algorithm based on the shortest capacity problem to compute this affinity function. We use these concept of meaning and semantic affinity to analyze and compare the gods and heroes from three different classical mythologies: Greek, Celtic and Nordic. We study the relationships of each individual mythology and those of common structure that is formed when we fuse the three of them. We show a strong connection between the Celtic and Nordic gods and that Greeks put more emphasis on heroic characters rather than deities. Our approach provides a technique to highlight and quantify important relationships in the original domain of the network not deducible from its structural properties. 
\end{abstract}



\section{Introduction}
One of the perpetual aims of human knowledge has been to unequivocally identify and represent the numerous concepts and sensations that populate human experience. Many philosophers and scientists from a wide range of scientific disciplines have given their own proposals in this topic, from physics \cite{disalle2006understanding} to philosophy \cite{friedman2013kant, heidegger2014introduction}, linguistics \cite{wittgenstein2013tractatus, wittgenstein2010philosophical} and computational science \cite{piccinini2010computation, pozuelo1969boundaries, ontological}.

Notable efforts have been devoted to model and compute with abstract concepts and ideas. Fuzzy logic, being the discipline devoted to work with imprecision and uncertainty, presents the paradigm of Computing with Words to express computations with words when there is not enough information to use numbers \cite{zadeh1999fuzzy, zadeh2013computing, zadeh1999computing}. Fuzzy Cognitive Maps have also been proposed, which are networks composed of different causal inferences that can be used to model complex phenomena \cite{papageorgiou2011learning, carvalho2013semantics, rickard2019modeling}.
In information retrieval semantic computing is used to infer the natural meaning of a concept \cite{wang2010formal, sheu2010semantic}. Computational ontologies model different ideas as a network and each pair of ideas is connected according to the relationship between them in a specific domain of knowledge \cite{guarino2009ontology}. Complex systems in applied linguistics model the words that humans used to denote concepts as a network of different interacting elements and their emergent properties naturally arise due to the actors interactions \cite{cameron2007complex, siebers2008introduction}. 


One emergent phenomena that has been widely studied due to its importance in human history is religion. Religious practices are as old as society. They are one of the pillars in which society and modern culture places their roots. 
Each civilization has had its own share of heroes, myths and gods that have helped shape the spirit and mind of the youth and the elder alike \cite{Stith1958TheBC}. Although nowadays some of the most obscure gods and traditions are only known to scholars, very popular heroes and gods are still alive in the collective consciousness of modern societies and they still echo in the characters present in many popular films and media \cite{reynolds1994super, campbell2008hero}. There is a long tradition in the study of the ancient myths and deities from a scientific point of view \cite{Frazer1922, capps1995religious,cirlot2004diccionario, elliott2014reinventing}.
The interest in these stories has not faded over time and is still a very researched topic in the humanities community \cite{liew2018present, lassen2019old, bengtson2016iarl}. The comparison and syncretism among the different gods that have populated the ancient mythologies has also been profoundly studied \cite{wen2020norse, keightley1877mythology, burkert1987oriental}. 
Besides the very-well known equivalence that bounds the Roman and Greek
gods, there are many other studies that show evidence of similarities between gods and myths from distant cultures \cite{Frazer1922, purzycki2013minds,  littleton1973new}. These relationships can be studied from a computational point of view considering, for example,
 agent-based social simulation \cite{dow2008} and social network analysis \cite{Chartash2019MathsMM, Yose2017ANS, NescolardeSelva2015, kenna2017networks, Miranda2018TheOM, Carron2012, Carron2013, akcca2021social, mac2012universal}.

Indeed, network science has become an important tool to study systems 
composed of interacting agents, such as proteins or human societies \cite{borgatti2009network, scott1988social, horvath2011weighted, Zhu2021May,barabasi2013network, idocin2020borgia, qazi2021using, wasserman1994social, cambria2016computational, Wild2021Feb}. 
One of the key ideas in social sciences is that living beings are embedded by our own social nature in a complex web of social relations and interactions \cite{Toth2021Feb, gong2016efficient}. This social fabric that we form has been traditionally modeled as a network, where each person is represented as a node that is connected to others according to 
some criteria. Social network analysis stands as an appropriate tool to understand many characteristics of the human behaviour, as it seems that many of us are deeply 
affected by the social structure in which we take part \cite{fischer1982dwell}. 
However, although the structural part of these networks has been carefully studied \cite{landherr2010critical}, there is no proper formalism to model each individual actor, taking into consideration its unique meaning and the relations that this meaning has with other actors. There is also no algorithm to quantify the resemblances of two concepts taking this concept into account.

In this paper we aim to build a new bridge between the philosophical issues that have been historically studied when defining ideas or concepts and the usual concepts of social network analysis to obtain new insights on different religious phenomena. We achieve these aims using different computational intelligence techniques: different operators very popular in fuzzy logic to compute affinity functions, which will determine the resulting semantic value for each actor in the network; and heuristic search algorithms, to compute it efficiently. We propose a new algorithm of this kind, the Pipe algorithm, designed to solve this task. Using these tools, we define the intrinsic, extrinsic and semantic value of a concept in a social network. Then, we show how these values can be used in the context of social network analysis as metrics to weight the importance of each concept, how the semantic value can be defined in terms of other concepts, and how we can perform interpretable semantic comparisons between one or more actors using the semantic value and affinity functions. By means of this proposal, we do not only give a computational framework to the topic of the ``Noumenon" or ``the-thing-in-itself" in Kantian philosophy \cite{rescher1981status}, but we also obtain novel results in comparative mythology.

The rest of the manuscript is organized as follows. Firstly, in Section \ref{sec:preliminars} we review some concepts regarding social network analysis and affinity functions. In Section \ref{sec:semantic_value} we present our formalization for the semantic value of an actor and how to use it as a centrality measure. Then, in Section \ref{sec:semantic_affinity_pipe} we present the semantic affinity concept and we detail the algorithm proposed to compute it. Subsequently, in Section \ref{sec:experiments} we show how we constructed a social network for each mythology, and the results obtained when analyzing each network using the semantic value and the semantic affinity, among other centrality measures. Finally, in Section  \ref{sec:conclusions} we discuss our results obtained, present some conclusions for this work, and establish the guidelines for our future research.

\section{Background} \label{sec:preliminars}
In this section we discuss some basic notions about centrality measures and affinity functions needed to understand how to model social networks and we explain affinity functions as a way to measure the relationship between a pair of actors.

\subsection{Centrality measures in social network analysis} \label{sec:preliminars_centrality}
In graph theory and network science, centrality measures indicate how relevant each node is in a structure \cite{landherr2010critical}. Centrality measures are important to identify the most important nodes in a network. Some very well known centrality measures are:

\begin{itemize}
	\item Degree centrality: the number of edges incident upon a node. In the case of directed networks, the degree is the sum of the number of edges incident to the node (in-degree) and the number of edges salient to the node (out-degree).
	\item Betweenness centrality: the betweenness of a node is the number of times that node is in the shortest path of other two nodes. It measures the brokerage ability of the node in the network's information flow.
	\item Closeness centrality: the closeness centrality of a node is the average length of the shortest path between that node and the rest of the nodes in the network. It measures the overall location of the node in the network, establishing a center-periphery difference.
	\item Eigenvector centrality: it assigns a relative score to each node in the network based on the idea that connections to well connected nodes should ponder more than connections to poorly connected ones.
\end{itemize}

\subsection{Affinity functions}
Affinity functions were defined in \cite{idocin2020borgia} as a way to measure the relationship between a pair of actors in a social network by capturing the nature of their local interactions. ``Affinities" are defined as functions over the set of actors of a given social network assigning a number between $0$ and $1$ to every pair of actors $x,y$: 
\begin{equation}
	F_C:(x, y) \rightarrow [0,1]
\end{equation}

Usually, $C$ is the adjacency matrix, each of whose entries $C(x, y)$ quantifies the strength of the relationship for the pair of actors $x, y$ in a weighted network, composed of $V$ different actors. The affinity between two actors shows how strongly they are connected according to different criteria, depending on which aspect of the relationship we are taking into account.

Affinities work similarly to a fuzzy set, substituting the idea of membership by the idea of affinity. So, a $0$ affinity value means that no affinity has been found at all while an $1$ value means that there is a perfect match according to the analyzed factors. For example, in the case of Best friend affinity, $0$ value means that there is absolutely no connection between the two actors and an $1$ value means that the the connection from $x$ to $y$ is the only connection that $x$ possesses. Since affinities are not necessarily symmetrical, the strength of this interaction depends on who the sender and receiver are, as it happens in human interactions e.g. unrequited love. So, it can happen that actor $x$ has an affinity of value $1$ with $y$, while $y$ has a much lower value with $x$.

In the following, we define some affinity functions (additional alternative definitions can be found in \cite{idocin2020borgia}):

\begin{itemize}
	\item Best friend affinity: it measures 
	the importance of a relationship with an agent $y$ for the agent $x$, in relation to all the other relationships of $x$: 
	\begin{equation} \label{eq-bf}
	F^{BF}_C (x,y)= \frac{C(x,y)}{\sum_{a \in V} C(x,a)}. 
	\end{equation}  
	
	\item Best common friend affinity: it measures the importance of the relationship taking into account how important are the common connections between the connected nodes to $x$ and $y$, in relation to all other relationships of $x$ in the network:
	\begin{equation} \label{eq-bcf}
	F^{BCF}_C (x,y)= \frac{\max_{a \in V }\{\min \{ C(x,a), C(y,a)\} \} }{\sum_{a \in V} C(x,a)}. 
	\end{equation}
	
	\item Machiavelli affinity: it computes how affine two actors $x$ and $y$ are based on how similar is the social structure that surrounds them:
	\begin{equation} \label{eq-mac}
	F^{Mach}_C (x,y)= 1 - \frac{\mid I_x - I_y \mid}{\max \{I_x, I_y\} }, 
	\end{equation} where $I_a = \sum_{z \in Z} D(z)$, where $Z$ is the set of actors where $C(a, z) > 0, \forall z \in Z$, and $D(z)$ is the centrality degree of $z$. 
\end{itemize}

There are two types of affinity functions: personal affinities and structural affinities. Personal affinities establish the strength of a interpersonal connection between $x$ and $y$ using  their respective connections and shared friends. Structural affinities quantify the relationship of a pair of actors based on the centrality measures of their nodes, such as their degree or betweenness. Two examples of personal affinities are the Best friend and Best common friend affinities, while the Machiavelli affinity is an example of a structural affinity.

Affinity functions have been used with different popular operators in fuzzy logic in order to generate new affinity functions using combinations of previously computed ones. Two of the most effective operators to combine different affinity functions are convex combinations of two affinity functions and T-norms of $n-$affinity functions. 

The idea of using convex combinations is that they ponder two different facets of a relationship, characterized by the two affinity functions to combine. T-norms are of particular importance because most affinity functions result in $0$ in a network, but some affinities, like the Machiavelli affinity, result in values bigger than $0$ most of the times. Using a T-norm is useful to maintain a low density in the resulting affinity network, because when one of the affinities is $0$ we know that the result of the aggregation will also be $0$.


\section{Semantic value in social network analysis}
\label{sec:semantic_value}
In this section we present our formalization of the semantic value of an actor in a network. The objective of this formalization is to model the ``Noumenon'' or ``the-thing-in-itself'' in Kantian philosophy \cite{rescher1981status}, and in many other famous philosophy works \cite{jacquette2007schopenhauer, cross1954logos}, in terms of computational concepts.

We define \textarc{\RR}, the semantic value of an actor in a social network, as the union of the intrinsic and the extrinsic value:

\begin{equation} \label{eq:semantic_full}
\textarc{\RR}(x)  = \cup (\textarc{\R}(x), \textarc{R}(x) )
\end{equation}

First we define \textarc{\R}($x$), the intrinsic value of an actor $x$ in a social network, as the property that is unique and inherent to him, not necessarily deducible from the network structure or topology. If actor $x$ was removed from the network, then all the present \textarc{\R}($x$) in the network is also removed from all the semantic values in the network.

Then, we define the extrinsic value of $x$, denoted as \textarc{R}($x$), as the property that represents the information of the local interactions of $x$, considering the relationships as a way of information or resource transmission, as usual in social network analysis \cite{wasserman1994social}. Being $X$ the vector of nodes connected to $x$, of dimension $a$, and $F_C$ an affinity function, the extrinsic value, $\textarc{R}(x)$, is a set defined as:

\begin{dmath} \label{eq:outer}
\textarc{R}(x) = \bigcup_{i=1}^{a} \{F_C(X_i, x)\textarc{\RR}(X_i) - \cup_{j\in J} \{ \cap ( F_C(X_i, x) \textarc{\RR}(X_i), F_C(X_j, x) \textarc{\RR}(X_j) ) \}
\end{dmath} 
where $J = \{j \in \{1, \dots, a\}, i \neq j\}$. With this expression, we stablish that the extrinsic value is the union of the received semantic value from the rest of the actors. Each actor in $X$ sends its own semantic value to $x$, modulated in each case by the affinity for that relationship, and erasing the redundancies from other relationships.

For the sake of simplicity, we can shorten Eq. (\ref{eq:outer}) by using $V_x(b) = F_C(X_b, x) \textarc{\RR}(X_b)$:

\begin{equation} \label{eq:outer_simple}
\textarc{R}(x) = \bigcup_{i=1}^{a} \{V_x(i) - \cup_{j\in J} \{ \cap ( V_x(i), V_x(j)) \}
\end{equation} 

This definition has one important issue: it has an infinite recursivity when computing the extrinsic value. Such recursivity is unavoidable for this definition and it is in line with the idea that dictionaries, in order to define a word, are required to use other words.

\subsection{Computing the semantic value}
\label{sec:methods_comp}
In order to give a computable version of the semantic value for a node $x$, it is first required to give a computable version of the intrinsic and extrinsic values, \textarc{\R}($x$) and \textarc{R}($x$). We denote the computable version of \textarc{\R} as $I$, the computable version of \textarc{R} as $E$, and the computable version of \textarc{\RR} as $S$.

Due to inherent fuzziness of the concept of intrinsic value, we cannot give an exact mathematical formula to compute it. However, depending on the context and the application, we can use a function to transform this abstract idea into a number. For example, if we are working with a network of routers, the real-world condition of each router location is an important part of each router \textarc{\R}. If we consider the fitness of each of these real-world conditions for the signal transmission task, we could use this fitness as $I$. As we always work with texts in the subsequent sections, we have approximated the intrinsic value of a node as the frequency (number of appearances) of the term associated to the node in the texts composing the original collection of documents. 

The computation of the extrinsic value is more complicated. First, it requires choosing an affinity function to quantify the relationships. 
Second, the extrinsic value has the said additional difficulty which is the endless recursivity that is present in Eq. (\ref{eq:outer}). To compute this, we have approximated the \textarc{\RR}($X_i$) in the original formula with $I(X_i)$, which eliminates the recursivity. Since we have numbers instead of sets in this case we use the summation instead of the union. The intersection of the received semantic value from $X_i$ and $X_j$ to $x$ is approximated as the result of propagating $I(X_i)$ through $X_j$ to $x$. So, the expression that approximates the extrinsic value is:
%

\begin{dmath}\label{eq:extrinsic_simplified_quant}
E(x) = \sum_{i=1}^{a} \max\left(F_C(X_i, x) I(X_i) - \sum_{j\in J}  F_C(X_i, X_j) I(X_i) F_C(X_j, x), 0\right)
\end{dmath}

where $J = \{j \in \{1, \dots, a\}, i \neq j\}$.

Finally, we can compute $S(x)$ with the analogous formula to Eq. (\ref{eq:semantic_full}):

\begin{equation} \label{eq:semantic_quant}
S(x)  = I(x) + E(x)
\end{equation}

\subsection{Semantic value as a centrality measure} \label{sec:semantic_measure}

Once we have computed $S$ for each node in a network, we can analyze the results just as with any other centrality measure. In order to have a high $S$, an actor $x$ must have a high $I$ and $E$ in comparison with the rest of the actors in the network.

To have a high $I$ value, $x$ must be important in the original domain from which we constructed the network. The function $I(x)$ does not tell much about the network structure directly but it can reinforce the importance of $x$ if $x$ has also high values in other centrality measures. If $x$ does not have high values in the remaining centrality measures but it has a high $I(x)$, this is revealing that $x$ was important in the original domain in a way that has not been taken into account when building the network.

To own a high $E$ value, the actor $x$ must have connections with other nodes with high $I$ that are not connected each with other. Low values of $E$ indicate that the actor relations are not very important in the network dynamics, it is part of a small community or its connections are few or redundant. 

For example, in the specific case of a word association network in a text, a domain-specific term is not likely to have very high semantic value. As it points out to a very narrow concept, it will not have a high $I$ value. Besides, since those domain-specific terms relate mostly to other specific terms of the same domain, they will not have many connections with other actors, so the possible extrinsic value is very limited. On the contrary, high semantic values indicate a very general concept. For example, it we take the word ``hand", it can be literally a hand, but it can also be used in many contexts, and it is closely related to the more general concept of ``utility". Someone can be ``the right hand of someone", something can ``get out of hand", or something can be ``at hand". The idioms regarding the ``hand" theme can be more general too: wash your hands means to give up the responsibility in a situation or can be interpreted literally, etc.

In order to compute the semantic value as a numerical value we approximate the intrinsic value as the frequency of the word in the original collection of texts. In the subsequent sections we denote the numerical version of the semantic value as $S$, the numerical version of the intrinsic value as $I$, and the numerical version of the extrinsic value as $E$. 
\section{Semantic affinity}
\label{sec:semantic_affinity_pipe}
The semantic affinity of two actors $x$ and $y$ measures the affinity between them based on the idea of how notably we need to change $S(x)$ to convert it to into $S(y)$. In such a way, terms that are similar in meaning should have high values of semantic affinity and non-related terms should have a very low semantic affinity. For example, the semantic affinity between ``water'' and ``ice'' should be high because these are very close terms and in real life we only need to freeze water below 0ºC to obtain ice. However, the semantic affinity between ``water'' and ``earth'' should be lower, as the difference in real life between those concepts is higher. We can compute the semantic affinity based on how efficient it is to propagate $S(x)$ into $S(y)$, using the Pipe algorithm, detailed in the next section.

\subsection{Pipe algorithm to compute the semantic affinity between two actors in a network} \label{sec:pipe}

The Pipe algorithm computes $A(x, y)$ based on the idea of modeling $S(x)$ as a liquid we need to carry from $x$ to $y$. Each actor $x$ has a capacity equal to its own semantic value $S(x)$ and each edge $x \rightarrow y$ can carry up to $FC(x,y) \cdot S(y)$ of that liquid. So, each edge is treated as a ``pipe" where the liquid goes and each actor as a bifurcation in the path. Then, we need to carry all the liquid from the source actor to the destination actor using the most efficient possible path.

There are different possibilities to define ``efficiency'' in this case. If we consider that efficiency means not requiring many actors to perform the transportation, we shall denote the most efficient path as that with less intermediate actors. If we consider that efficiency means using the minimum capacity path, then we shall consider the optimal path as the one that used the intermediate actors with the lowest semantic values. We can also define efficiency as using only good connections, in that case the optimal path is that where the average affinity value is the highest. In this work we have opted for the latter formulation and we have denoted the most efficient path as the one with the highest average affinity value. 

In order to compute the most efficient path we need to take the capacity of each actor and the affinity in each edge into account. This problem is quite similar to the Shortest Capacitated Path Problem \cite{costa2002bounds}, which consists of finding a set of edge-disjoint paths that connects all the nodes in a graph, but in our case we are only taking one path into account, from $x$ to $y$. The classical ``shortest path'' optimization problem between a pair of nodes \cite{goldberg2005computing} is also very related to the desired task. However, this problem solves the shortest path and we are looking for the most efficient one. We can reformulate the problem by rescaling the affinity values so that the shortest path is actually the one with the highest average affinity value.

In this way, we obtain the shortest possible path using all the nodes and edges that are not yet ``full". Usually, it is required to use more than one path to carry all the semantic value from the source to the destination. So, we have to compute a new shortest path with the available nodes and edges every time the actual path has already met its capacity limit. Once we have carried all the liquid from one node to another, we compute the final result taking into account the original difference in their semantic values $S$ and the average affinity value in the edges used to carry it. 

Since some actors naturally have low affinity values, for example when they have a lot of connections, the expected value of the average affinity values of a path can be deceptively low. In order to better compare the different semantic affinities that originate from actor $x$, we rescale the result by the maximum semantic affinity that $x$ emits. This leads us to the final expression of $A(x,y)$:
\begin{dmath}
A(x, y) = \left(1 - \frac{|S(x) - S(y)|}{\max(S(x), S(y))}\right) \frac{\sum P}{|P|} \cdot \frac{1}{ \lor_{z\in Z} F_C(x, z)}
\end{dmath}

\noindent where $P$ is the list of affinity values in the edges used in the paths to carry the semantic value, $S(x)$, from $x$ to $y$, and $Z$ is the set of all actors connected to $x$.

\begin{figure*}[h]
	\centering
	\captionsetup[subfloat]{labelformat=empty}
	\subfloat[width=.3\linewidth][\textbf{1.} We start with ``Apollo'' containing his full semantic value, that will be propagated to ''Zeus'']{\includegraphics[width=.31\linewidth]{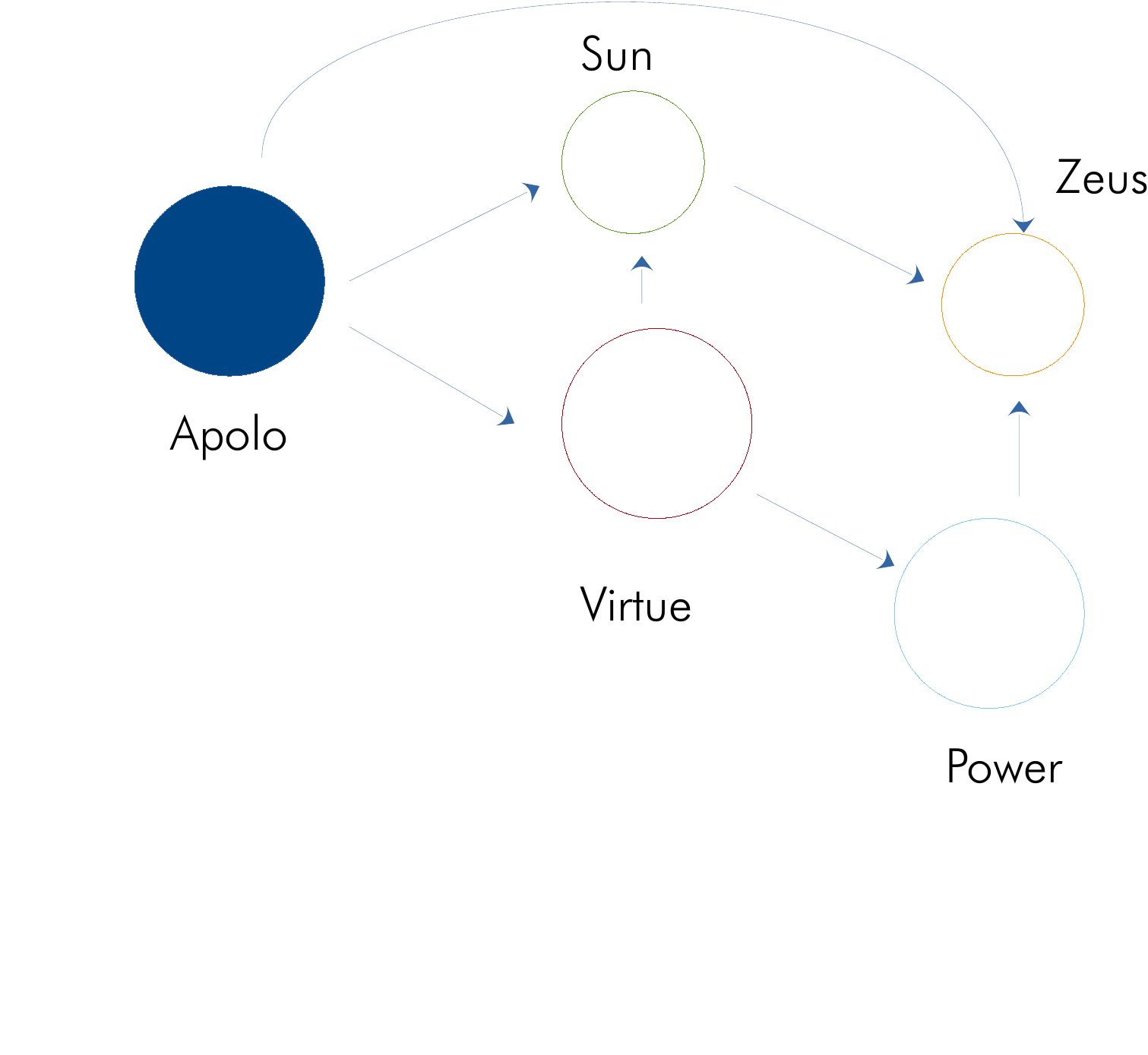}}\hspace{.1cm}
	\subfloat[width=.3\linewidth][\textbf{2.} First, the semantic value is propagated using the affinity function between the two actors. Since it is not enough to carry all the semantic value, we proceed to look for another path.]{\includegraphics[width=.31\linewidth]{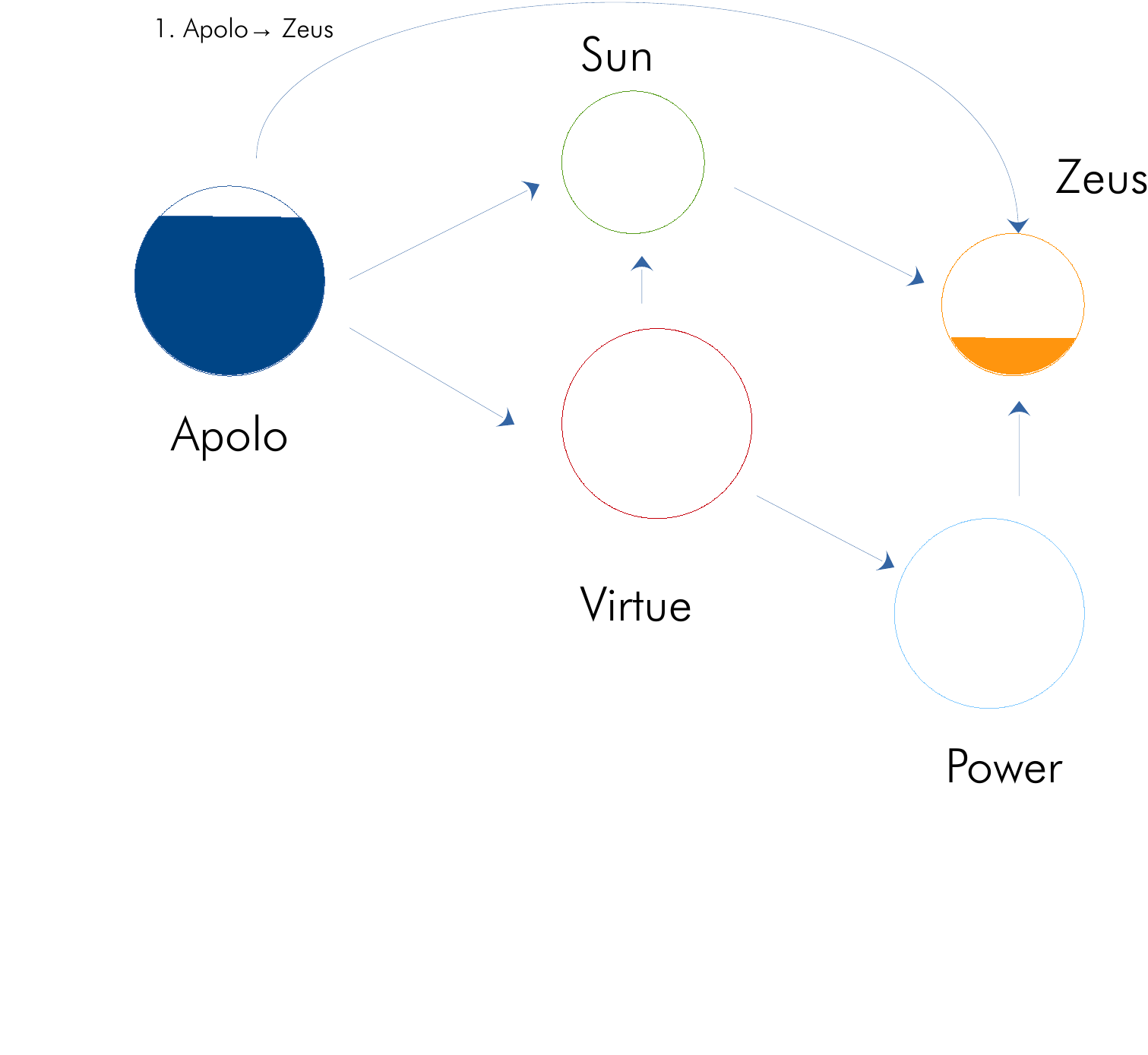}}\hspace{.1cm}
	\subfloat[width=.3\linewidth][\textbf{3.} We propagate all the possible semantic value thorugh ``Sun'', and then to ``Zeus''. We have saturated the capacity of this path, but not the capacity of `Sun''.]{\includegraphics[width=.31\linewidth]{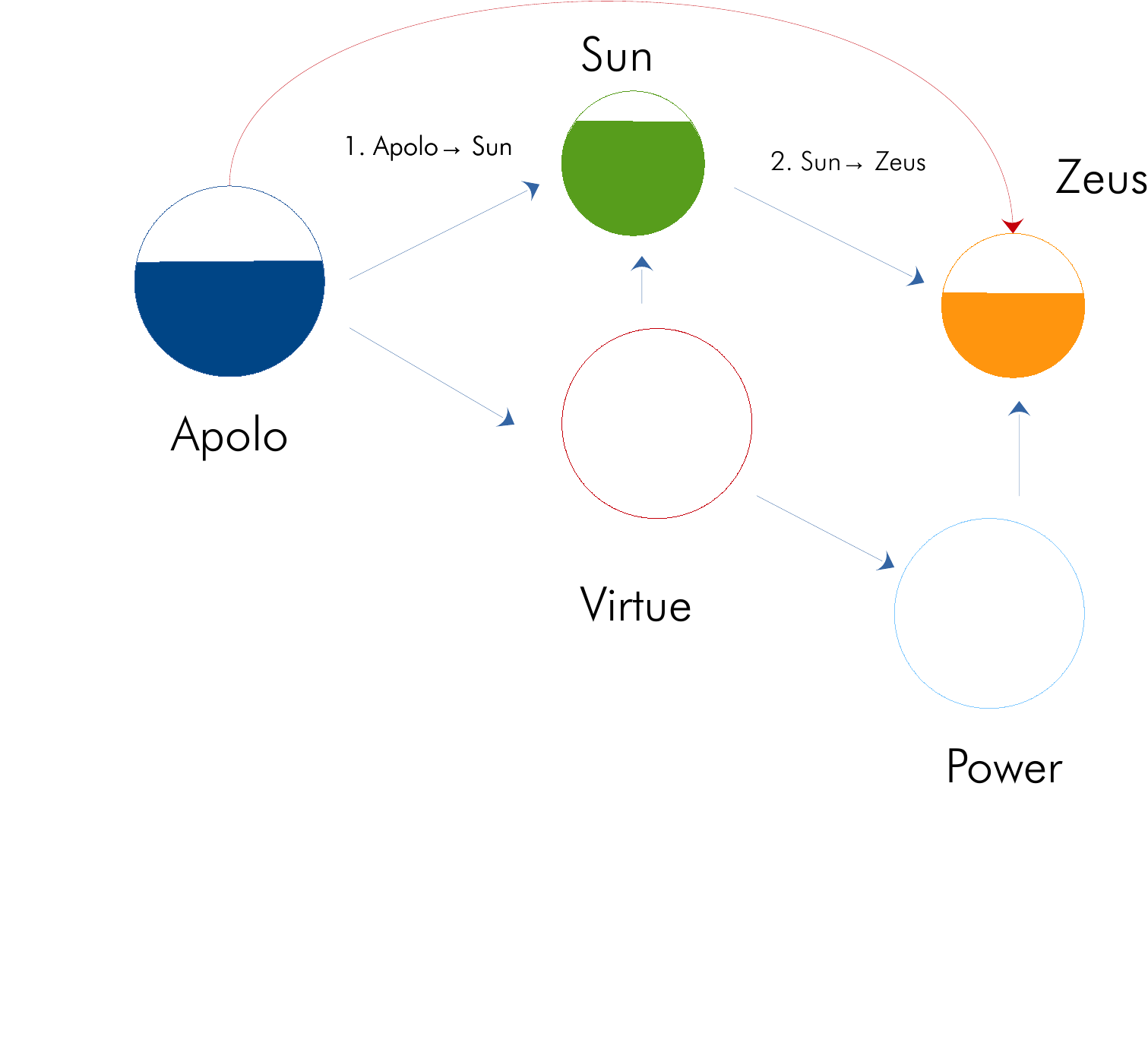}}\\
	\subfloat[width=.3\linewidth][\textbf{4.} We use the affinity between ``Apollo'' and ``Virtue'', and between ``Virtue'' and ``Sun'' to fill the rest of ``Sun''. We have filled ``Sun'' completely, but we can still propagate thorugh ``Virtue''.]{\includegraphics[width=.33\linewidth]{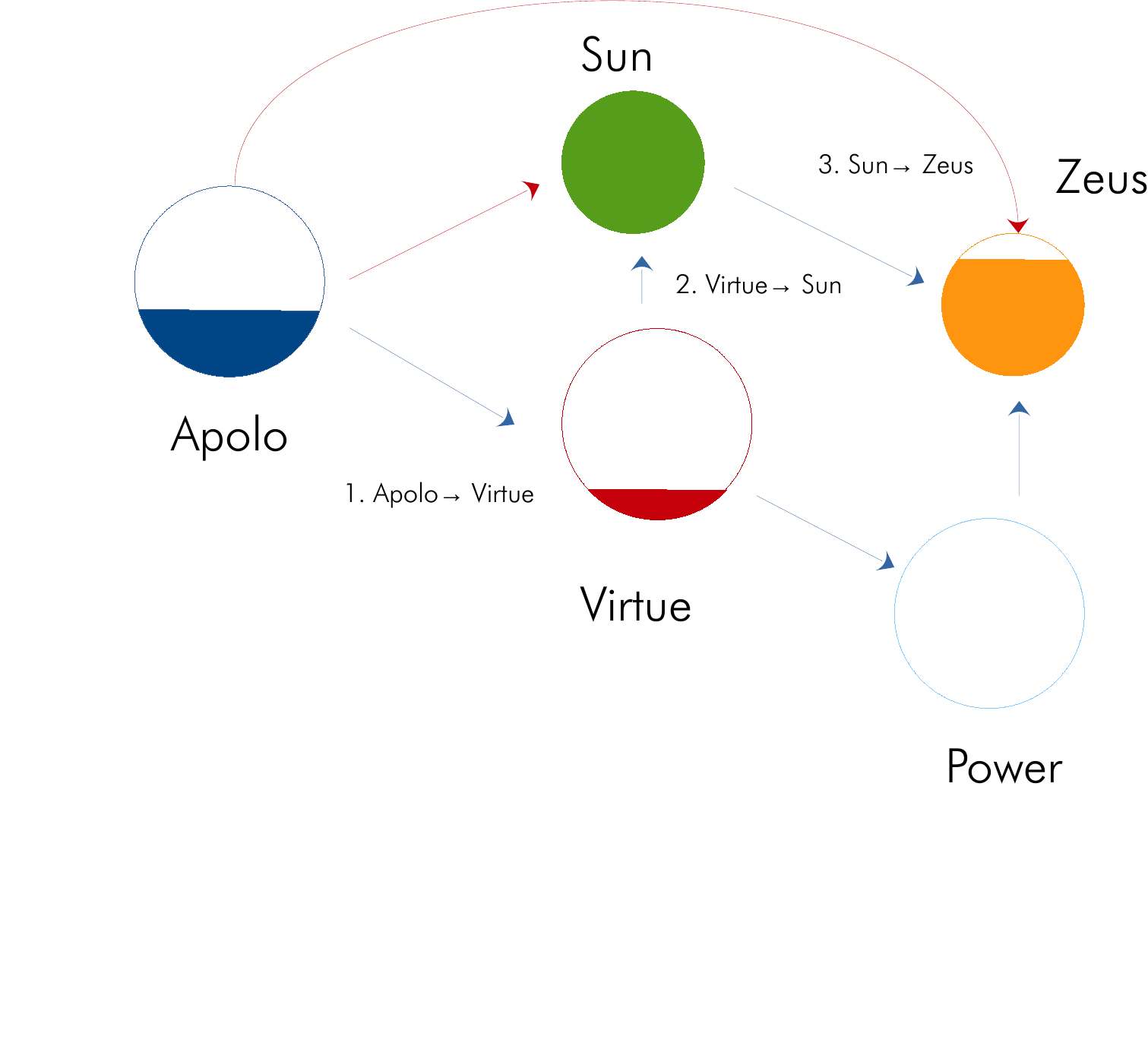}}\hspace{.3cm}
	\subfloat[width=.3\linewidth][\textbf{5.} We fill ``Zeus'' using the affinity between ``Apollo'' and ``Virtue'', ``Virtue'' and ``Power'' and ``Power'' to ``Zeus''.  Some semantic value is still in Apollo, but ``Zeus'' is already filled so we cannot propagate more.]{\includegraphics[width=.33\linewidth]{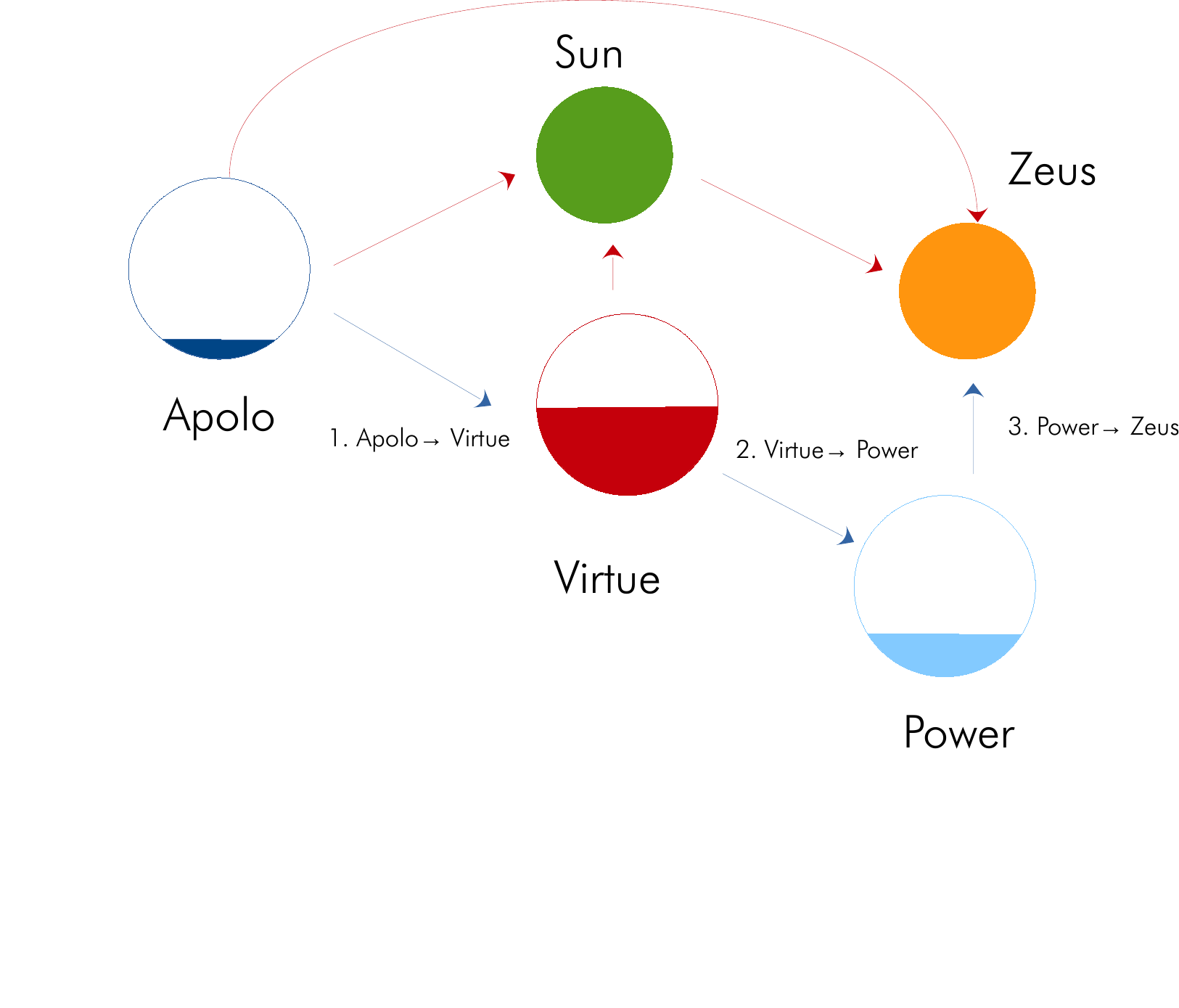}}
	\caption{\textbf{Example for a simplified execution of the Pipe algorithm to compute the semantic affinity between ``Apollo'' and ``Zeus''}. Each subfigure describes the multiples steps taken to carry the semantic value from one actor to another. In this case we were able to fill ``Zeus'' completely, but we needed to use all the possible affinity values in the network, and some semantic value from ``Apollo'' did not fit in ``Zeus''.}
	\label{fig_pipe_scheme}
\end{figure*}

The pseudo-code for the pipe comparison algorithm is in Algorithm \ref{alg:pipe} and in Figure \ref{fig_pipe_scheme} we show a graphical example of one execution of this algorithm.

In order to compute the semantic affinity in our experimentation, we have used a combination of best friend and Machiavelli affinities as the $F_C(x, y)$. Using this mix of affinity functions we can characterize each edge based on the importance of the pairwise relationship between $x$ and $y$, and also take into account the relative importance that both of their social circles have in the network. This is important for two reasons: 
\begin{enumerate}
	\item In high degree actors, the best friend affinity values are necessarily low, which will result in artificially low semantic values.
	\item The Machiavelli affinity gives high affinity to actors that play a similar role in the network. In the texts we are studying, these results in high affinity values to concepts that play a similar role in the tales.
\end{enumerate}
So, it is natural to think that those actors are transmitting information between them, even though the best friend affinity value between them is not high. We have combined both affinity functions using a convex combination, so the value of each edge is $90\%$ the best friend affinity value and $10\%$ the Machiavelli affinity, but we set to $0$ all affinity values in the edges where the original Best friend affinity was $0$.

\begin{algorithm}
	\caption{Pipe Comparison Algorithm}\label{alg:pipe}
	
	\begin{algorithmic}[]
		\Function{fillPath}{$c$, $x$, $y$, $edgesSeen$, $F_C$, $M$}
		\If {$M(c(0)) \le 0$}
		\State
		\Return []
		\EndIf
		
		\State $lenPath = length(c) - 1$
		\State $liquidCarried \leftarrow 0$
		\State $restingLiquid \leftarrow M(x)$
		
		\For {$i \in range(lenPath)$}
		\State $afPath \leftarrow F_C(c(i), c(i+1))$
		\State $carry \leftarrow min(m(c(i+1), afPath*restingLiquid)$
		\State $M(c(i+1)) \leftarrow max(0, capacities(c(i+1)) - afPath*carry)$ 
		\State $liquidCarried \leftarrow liquidCarried + carry$
		\State $append(edgesSeen, ((c(i), c(i+1)))$
		\EndFor
		\State $M(c(0)) \leftarrow M(c(0)) -  liquidCarried$
		\State 
		\Return $edgesSeen$
		\EndFunction
	\end{algorithmic}
	
	\begin{algorithmic}[]
		\Function{PipeComparison}{$G$, $x$ , $y$}
		\State $F_C \leftarrow buildAffinityNetwork(G)$
		\State $M \leftarrow computeSemantics(G)$
		\State $shortestPaths \leftarrow allEfficientPaths(F_C, x, y)$
		\State $edgesSeen \leftarrow []$
		\For{$c \in allEfficientPaths$}
		\State $append(edgesSeen,$ \\$ fillPath(c, x, y, edgesSeen, F_C, M))$
		\EndFor
		\State 
		\Return $(1 -|M(x) - M(y)| / \max(M(x), M(y))) * average(edgesSeen) / \max(Af(x, :))$
		\EndFunction
	\end{algorithmic}
\end{algorithm}

\section{Comparative mythology analysis using the semantic value and the semantic affinity}
\label{sec:experiments}

In this section we discuss the different steps to perform our mythology analysis:
\begin{enumerate}
	\item How we constructed the network for each mythology, and how we fused them.
	\item The results for the centrality measures and the semantic value in each network.
	\item The affinity values for the semantic, best friend and Machiavelli affinities for important characters in their respective mythologies.
\end{enumerate}

\subsection{Building the mythology networks} 

In this section we show how we built the network for each mythology. We discuss which books were used to form each network, some statistics regarding word counts, and how we processed the text to obtain the desired networks.

\subsubsection{Processing the texts}\label{sec:fusion}
We have chosen three of the ancient mythologies to perform the comparative study: Greek, Nordic, and Celtic. We have opted for these three due to their well-known interest and the existence of available compilations of tales translated to English, which makes the text processing for each book much easier. We have chosen selected the following books as a basis for our analysis:

\begin{itemize}
	\item \emph{Celtic Wonder-Tales} by Ella Young (1867-1956) \cite{young1923celtic}. Originally written in 1910, it is a collection of Celtic traditional tales translated to modern English.
	\item \emph{Greek Myths} by Olivia Collidge (1908-2006) \cite{olivia1948}. Is a compilation of various stories regarding the classical Greek pantheon in modern English.
	\item \emph{The Younger Edda} by Snorri Sturluson (1179-1241) \cite{edda1916}. \emph{The Prose Edda} or \emph{The Younger Edda} is a medieval Icelandic compilation of mythical texts, made by Snorri Sturluson, who was a historian, politician and poet in Iceland \cite{wanner2008snorri}. The degree of originality that he added to this compilation is unclear but the original stories contain material from traditional sources, reaching the Viking Age. 
\end{itemize}

Since we have the plain text files, it is easy to extract each chapter/tale in each book. We then parse each of them following the standard procedure \cite{webster1992tokenization, fox1992lexical} using a pre-trained multilayer perceptron in the Python Natural Language ToolKit \cite{bird2009natural}. We purge every word that is not a noun, since we only want to model interaction between entities and concepts. In Table \ref{tab:book_counts} we report the size of each book and the number of entities found.
All the texts have been obtained from the Gutenberg Project \cite{stroube2003literary}.

\begin{table}
	\centering
	\caption{\textbf{Report of the size of each mythology.} Number of words, chapters and entities for each book in this work.}
	\adjustbox{max width=\linewidth}{
	\begin{tabular}{ccccc}
		\toprule
		Mythology & Book & Chapters & Words & Entities \\
		\midrule
		Celt			  & \emph{Celtic Wonder-Tales}& 13 & 41613 & 5114 \\
		\midrule
		Greek & \emph{Greek Myths} & 27 & 61246 & 5985 \\
		\midrule
		Nordic & \emph{The Younger Edda} &  21 & 65388  & 7521\\
		
		\bottomrule
	\end{tabular}
}
	
	\label{tab:book_counts}
\end{table}

\subsubsection{Obtaining the networks}

Once we have extracted the nouns from the text, what we have is a series of stemmed tokens. To obtain a network, we need the nodes and the edges to form it. In the case of the nodes, we will make a bijective association, so that one noun will correspond to one node, and {\it vice versa}. There are different ways to compute the edges in terms of noun co-occurrence. We have decided to create an edge every time a word appears in a $k$-distance or less from another in the text, choosing $k$ as 10.


\subsubsection{Fusing the networks}
\begin{table}
	\centering
	\caption{\textbf{Entity intersection for the three cultures.} Percentage of entities in each culture that also are present in another one.}
	\begin{tabular}{lccc}
		\toprule
		Common entities & Celt & Greek & Nordic \\
		\midrule
		Celt & - & 19.59\% & 17.20\%\\
		Greek & 19.59\% & - & 23.08\%\\
		Nordic & 17.20\%& 16.69\% & -\\
		\bottomrule
	\end{tabular}
	
	\label{tab:intersection}
	
\end{table}

Given the network for each tale, we can fuse them to obtain a ``global" network containing the information from all the different networks referring to each tale. Since many of these stories share a fair group of topics, they all have a significant amount of common terms with the others (Table \ref{tab:intersection}). There are no problems of scale in this context, since all stories range from 2 to 7 pages long only. So, we simply add up all the edges into a single network. When an edge between two actors is repeated in various of the networks, we take the highest value.

\subsection{Analysis of the semantic value and centrality measures in the myth networks}
\begin{figure*}
	\centering
	\subfloat[]{\includegraphics[width=0.45\linewidth]{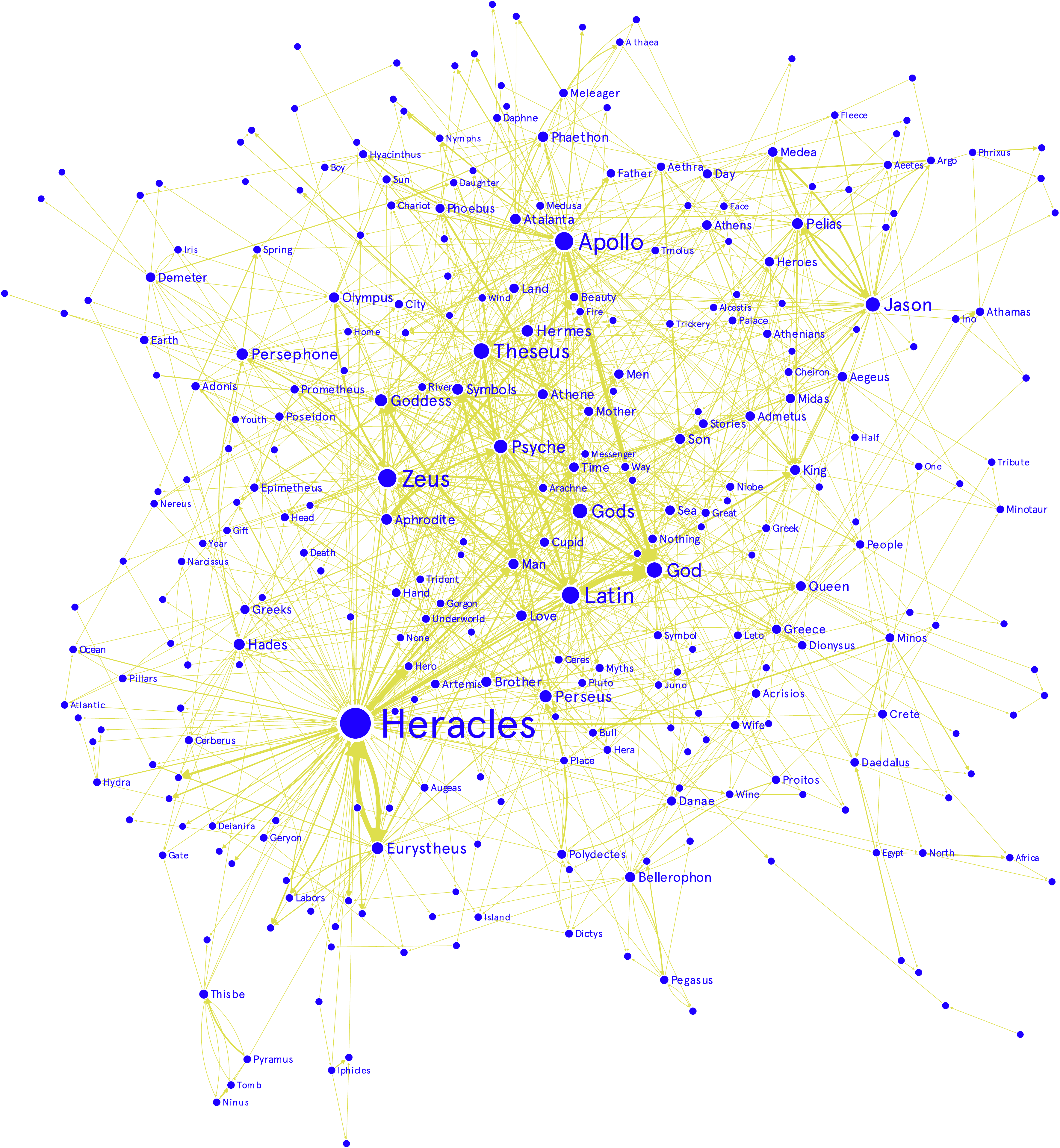}}
	\subfloat[]{\includegraphics[width=0.45\linewidth]{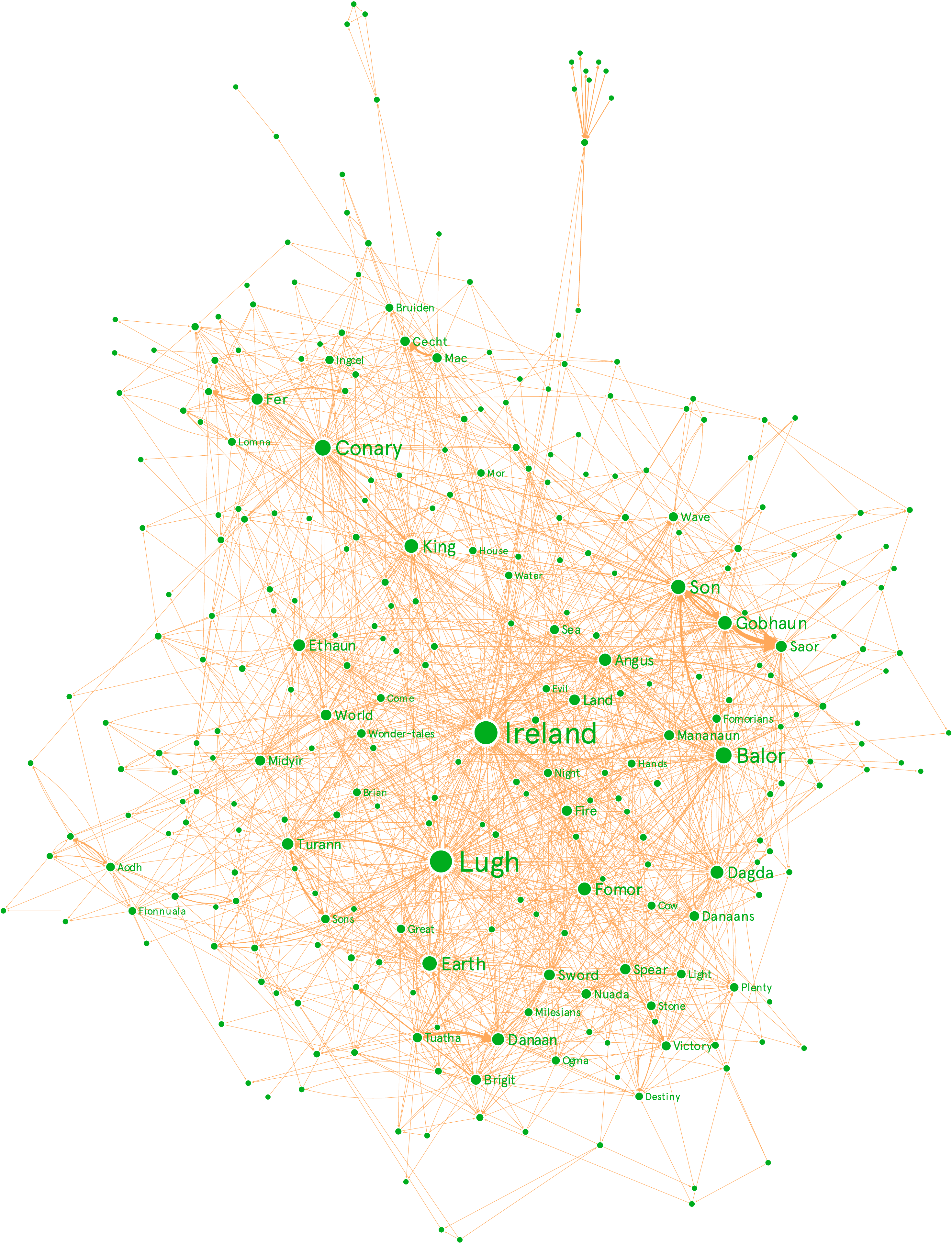}}
	\\
	\subfloat[]{\includegraphics[width=0.45\linewidth]{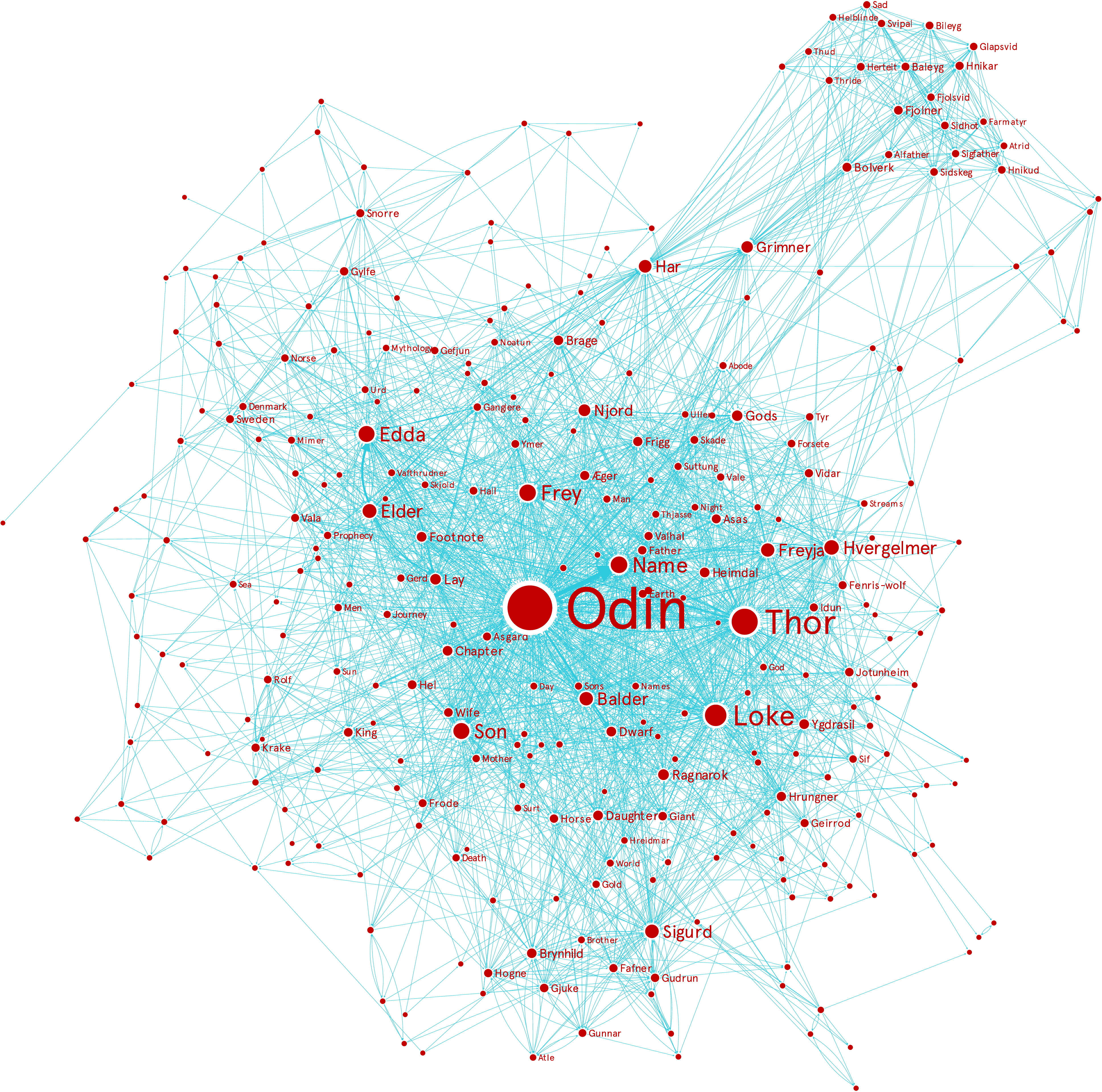}}
	\subfloat[]{\includegraphics[width=0.45\linewidth]{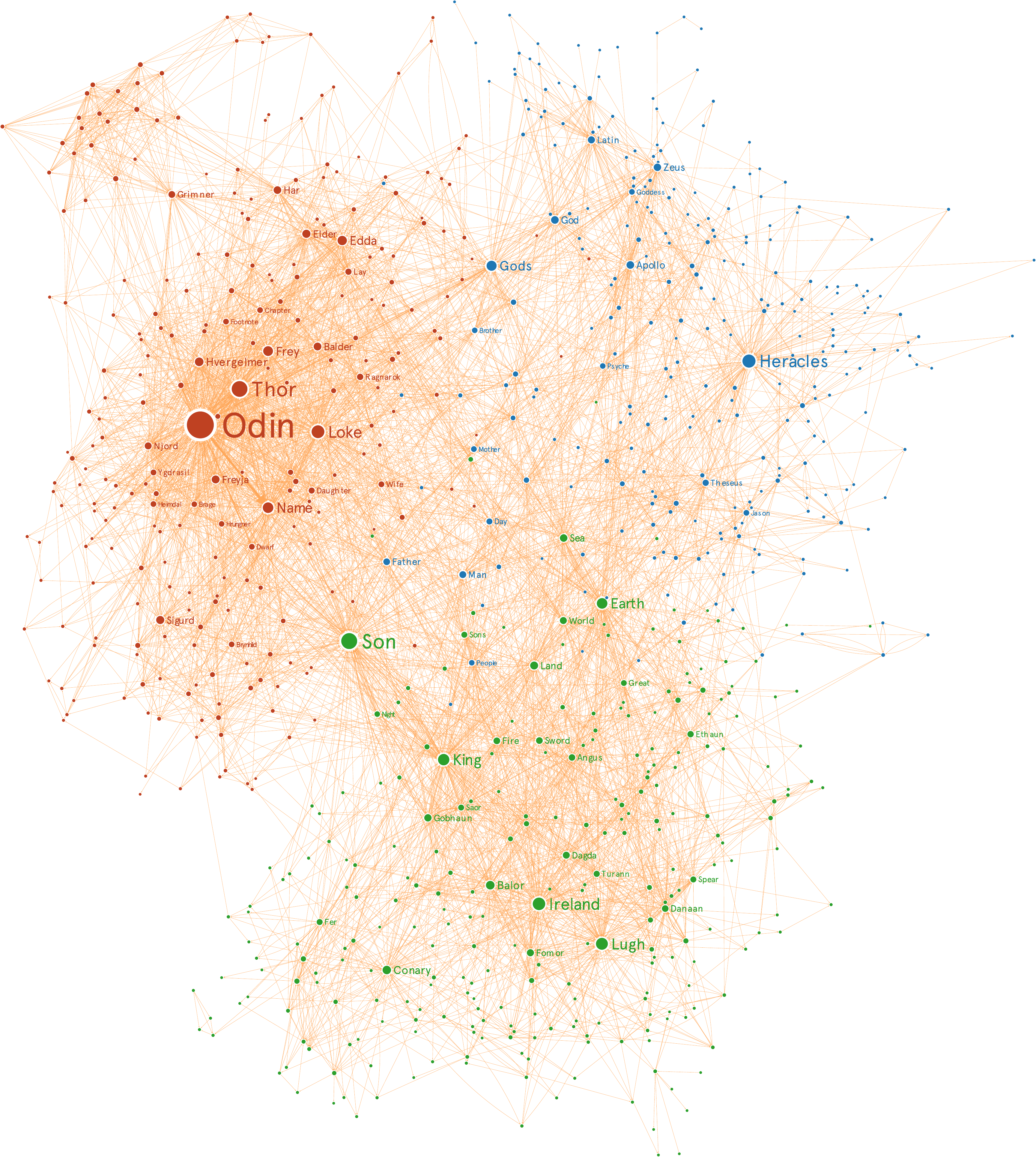}}
	\caption{\textbf{Word co-occurrence networks}. Each network is formed using the 300 most repeated entities in each corpus. We consider a connection between two words every time they appear less than 10 words apart from each other in one of the texts analyzed. \textbf{a.} \emph{Greek Myths} \textbf{b.} \emph{Celtic Wonder-Tales} \textbf{c.} \emph{Younger Edda} \textbf{d.} Fusion network of the three cultures. For the fusion network, color is attributed to each node according to the frequency in each corpus. Red means majority of appearances in \emph{Younger Edda}, blue in \emph{Greek tales} and green in \emph{Celtic Wonder-Tales}. Node size is directly proportional to the in-degree measure and the layout algorithm considered is Force Atlas 2 \cite{jacomy2014forceatlas2}. }
	\label{fig:networks}
\end{figure*}

\begin{table}
	\centering
	\caption{\textbf{Centrality measures in \emph{Greek Myths} network}. For the top 10 most repeated entities in the associated texts.}
	\subfloat{\adjustbox{valign=b, width=\linewidth}{\begin{tabular}{lccccccc}
				\toprule
				{} &    $S$ &   $E$ &  Freq. ($I$) &  Degree &  Betweenness &  Closeness &  Eigencentrality \\
				\midrule
				Heracles   & 239.49 & 89.49 &          150 &     364 &         0.48 &       0.52 &             0.53 \\
				Theseus    &  84.69 & 29.69 &           55 &     118 &         0.10 &       0.44 &             0.21 \\
				King       &  84.15 & 27.15 &           57 &      55 &         0.03 &       0.42 &             0.13 \\
				Jason      &  76.93 & 27.93 &           49 &      88 &         0.06 &       0.37 &             0.13 \\
				Apollo     &  64.83 & 20.83 &           44 &      86 &         0.08 &       0.40 &             0.13 \\
				Psyche     &  64.72 & 15.72 &           49 &      91 &         0.07 &       0.39 &             0.13 \\
				Eurystheus &  57.54 & 25.54 &           32 &      77 &         0.01 &       0.39 &             0.16 \\
				Zeus       &  57.27 & 21.27 &           36 &      75 &         0.09 &       0.43 &             0.14 \\
				Perseus    &  45.30 & 16.30 &           29 &      54 &         0.04 &       0.36 &             0.08 \\
				Pelias     &  43.76 & 21.76 &           22 &      48 &         0.02 &       0.37 &             0.09 \\
				\bottomrule
		\end{tabular}}
	}
	
	\label{tab:greek}
\end{table}
\begin{table}
	\centering
	\caption{\textbf{Centrality measures in \emph{Celtic Wonder-Tales} network}. For the top 10 most repeated entities in the associated texts.}
	\adjustbox{valign=b, width=\linewidth}{
		\begin{tabular}{lccccccc}
			\toprule
			{} &    $S$ &   $E$ &  Freq. ($I$) &  Degree &  Betweenness &  Closeness &  Eigencentrality \\
			\midrule
			Lugh    & 130.73 & 55.73 &           75 &     258 &         0.13 &       0.51 &             0.27 \\
			Ireland & 107.93 & 56.93 &           51 &     256 &         0.24 &       0.56 &             0.32 \\
			Conary  & 101.44 & 48.44 &           53 &     198 &         0.11 &       0.49 &             0.13 \\
			King    &  79.73 & 40.73 &           39 &     133 &         0.07 &       0.50 &             0.17 \\
			Son     &  78.22 & 41.22 &           37 &     176 &         0.06 &       0.49 &             0.18 \\
			Balor   &  73.10 & 38.10 &           35 &     186 &         0.08 &       0.49 &             0.20 \\
			Gobhaun &  71.95 & 23.95 &           48 &     147 &         0.04 &       0.46 &             0.17 \\
			Ethaun  &  63.92 & 27.92 &           36 &     100 &         0.04 &       0.44 &             0.09 \\
			Fomor   &  56.07 & 24.07 &           32 &     100 &         0.03 &       0.47 &             0.17 \\
			Turann  &  54.17 & 22.17 &           32 &     105 &         0.03 &       0.44 &             0.11 \\
			\bottomrule
		\end{tabular}
	}
	
	\label{tab:celtic}
\end{table}

\begin{table}
	\centering
	\caption{\textbf{Centrality measures in \emph{The Younger Edda} network.} For the top 10 most repeated entities in the associated texts.}
	\adjustbox{valign=b, width=\linewidth}{
		\begin{tabular}{lccccccc}
			\toprule
			{} &    $S$ &    $E$ &  Freq. ($I$) &  Degree &  Betweenness &  Closeness &  Eigencentrality \\
			\midrule
			Odin      & 215.00 & 106.00 &          109 &    1113 &         0.47 &       0.75 &             0.37 \\
			Thor      & 176.63 &  44.63 &          132 &     508 &         0.14 &       0.61 &             0.26 \\
			Loki      & 100.58 &  34.58 &           66 &     291 &         0.06 &       0.56 &             0.22 \\
			King      &  53.63 &  13.63 &           40 &      79 &         0.02 &       0.49 &             0.07 \\
			Frey      &  51.01 &  20.01 &           31 &     167 &         0.02 &       0.53 &             0.17 \\
			Har       &  50.24 &  16.24 &           34 &     116 &         0.03 &       0.52 &             0.09 \\
			Sigurd    &  45.02 &  19.02 &           26 &     178 &         0.02 &       0.52 &             0.13 \\
			Balder    &  43.69 &  14.69 &           29 &     151 &         0.02 &       0.52 &             0.14 \\
			Freyja    &  28.78 &  10.78 &           18 &     154 &         0.01 &       0.51 &             0.16 \\
			Norse     &  24.06 &   4.06 &           20 &      37 &         0.00 &       0.46 &             0.04 \\
			\bottomrule
		\end{tabular}
	}
	\label{tab:nordic}
\end{table}

\begin{table}
	\centering
	\caption{\textbf{Centrality measures in the fusion network for the 10 most repeated entities in every text analyzed.}}
	\adjustbox{valign=b, width=\linewidth}{
		\begin{tabular}{lccccccc}
			\toprule
			{} &    $S$ &    $E$ &  Freq. ($I$) &  Degree &  Betweenness &  Closeness &  Eigencentrality \\
			\midrule
			Heracles & 244.33 &  94.33 &          150 &     368 &         0.15 &       0.43 &             0.07 \\
			Odin     & 227.31 & 118.31 &          109 &    1121 &         0.17 &       0.48 &             0.33 \\
			King     & 222.12 &  86.12 &          136 &     269 &         0.09 &       0.48 &             0.13 \\
			Thor     & 178.63 &  46.63 &          132 &     502 &         0.05 &       0.44 &             0.24 \\
			Lugh     & 131.88 &  56.88 &           75 &     259 &         0.04 &       0.41 &             0.08 \\
			Son      & 116.97 &  77.97 &           39 &     414 &         0.11 &       0.49 &             0.22 \\
			Ireland  & 116.43 &  65.43 &           51 &     261 &         0.08 &       0.43 &             0.10 \\
			Conary   & 104.54 &  51.54 &           53 &     196 &         0.03 &       0.40 &             0.05 \\
			Loki     & 103.06 &  37.06 &           66 &     293 &         0.02 &       0.43 &             0.20 \\
			Theseus  &  80.19 &  25.19 &           55 &     116 &         0.02 &       0.40 &             0.04 \\
			\bottomrule
		\end{tabular}
	}

%
	\label{tab:myth}
\end{table}

We computed the word association networks for each of the mythologies studied and then we fused the networks into a single one, which are shown in Figure \ref{fig:networks}. Then, we computed the intrinsic, extrinsic and semantic value for each actor. We also computed other common centrality measures social network analysis \cite{newman2018networks}. We showed these results for the top most important actors according to semantic value in Tables \ref{tab:greek}, \ref{tab:celtic}, \ref{tab:nordic} and \ref{tab:myth}.

We have found that ``Heracles'' is the most important actor in the \emph{Greek Myths} network, according to all the measures taken. There are other important heroes in this list like ``Theseus'', ``Jason'', and ``Perseus''. All of them are somewhat the embodiment of bravery and authority, so it is not surprising that ``King'' has also a high semantic value. There are more human characters than gods: ``Apollo'' and ``Zeus'' are the only ones which appear at the top, with similar $S$ values, but not as high as the other Greek heroes here present. Regarding the classical centrality measures studied, the betweenness gives the highest value to ``Hercules'' by a large margin and penalizes specially ``King'' compared to the other metrics. The closeness does not show such a big gap between ``Hercules'' and the other actors, and similarly to the betweenness prefers ``Zeus'' over the human actors that posses more semantic value than him. This also happens in a smaller scale in the eigencentrality, that also preferred ``Eurystheus'' over the rest of the heroes. In general terms, classic centrality measures preferred gods, while the semantic value valued human and heroic figures  (Table \ref{tab:greek}).

In the case of \emph{Celtic Wonder-Tales}, we found ``Lugh'', the most prominent god of the Irish pantheon, to own the highest $S$ value followed by ``Ireland''.  The third actor in $S$ value, ``Conary'', is an important mythical king of Ireland whose reign ends when he breaks three sacred oaths. The concept of ``King'' also has a high $S$, just as in the \emph{Greek tales} case. Regarding the classical centrality measures, all of the them rated most highly ``Ireland''. Betweenness and closeness seem to be quite correlated in this case, showing the same top 3, but the betweenness quickly decreases after that. The eigencentrality values ``Balor'' significantly compared to the other metrics computed as it ranks in the third position. Contrary to most classical centrality measures, the semantic value favored the mythical embodiments of kingship like ``Lugh'', ``Conary'', ``Balor'', and ``Ethaun''. This fact, alongside the high $S$ value of ``Ireland'', indicates a strong connection in this compilation between these mythical figures and the sovereign of the country (Table \ref{tab:celtic}). Such bond was not found in the other two mythologies.

In \emph{The Younger Edda} we found ``Odin'', one of the main gods of the Germanic pantheon, to be the  most important actor in terms of $S$. Being the father of all the \AE sir, but also wise in the ways of magic and divination, the strength of these two different attributions might be the origin of such high $S$ value. Following ``Odin'', there is ``Thor'', another character with many attributions in his tales. A total of 6 gods populate this ranking, which shows that the gods themselves are more important in this mythology than in the other two. Regarding the four classical centrality measures, the top 3 is the same than in the semantic value. After the top 3, all the metrics prefer characters rather than concepts, but the semantic value puts ``king'' as the top 4 value, in consonance with the other compilations, which is an important difference (Table \ref{tab:nordic}).

Finally, for the case of the fused network, we can see how the three fused structures can be recognized in the final structure but also numerous bridges have appeared to join them. As expected, these bridges are mostly general concepts, such as ``Son'', ``Gods'' and ``Father'', which connect the specific deities for each mythology. When analyzing the semantic values, ``Odin'' is again the one with the highest value. However, in this case the correlation of $I$ and $S$ seems to be less important. For example, ``Son'', with an $S=78.12$, did not enter in the top 10 most frequent words, neither ``Earth'' with $S=44.63$. Comparing the semantic value to the remaining classic centrality measures, ``Odin'' generally gets the top value for them, which again reinforces the tendency of the the semantic value to prefer human and heroic figures over deities. The eigencentrality and the degree centrality generally favored actors from the \emph{Younger Edda}, probably because they form a more densely connected structure. The closeness put ``Son'' as the top value, since `Son'' is an important bridge between the Nordic and Celtic entities. 

\subsection{Semantic affinity analysis in the myth networks}

\begin{figure}
	\centering
	\subfloat[]{\includegraphics[width=0.49\linewidth]{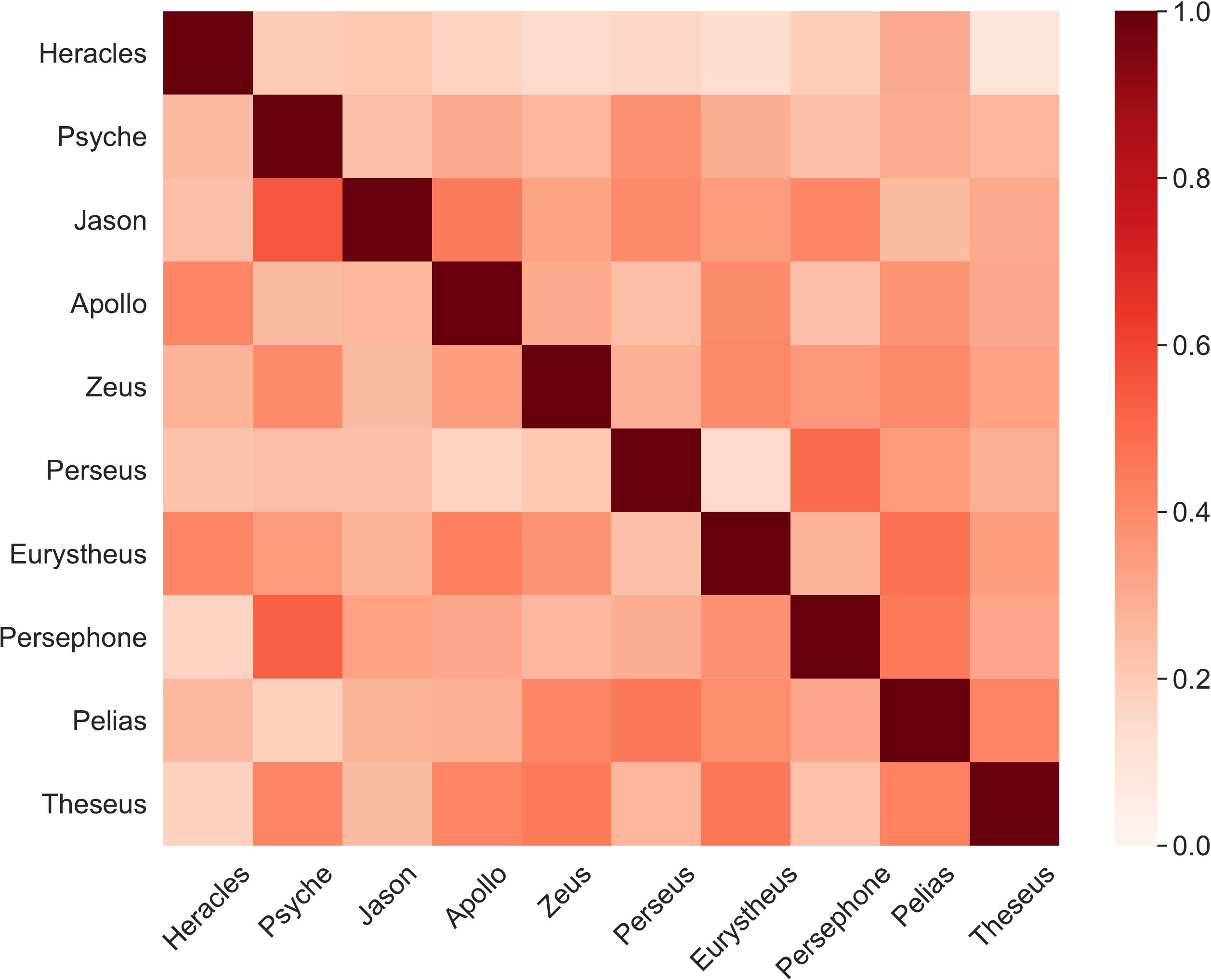}}
	\subfloat[]{\includegraphics[width=0.49\linewidth]{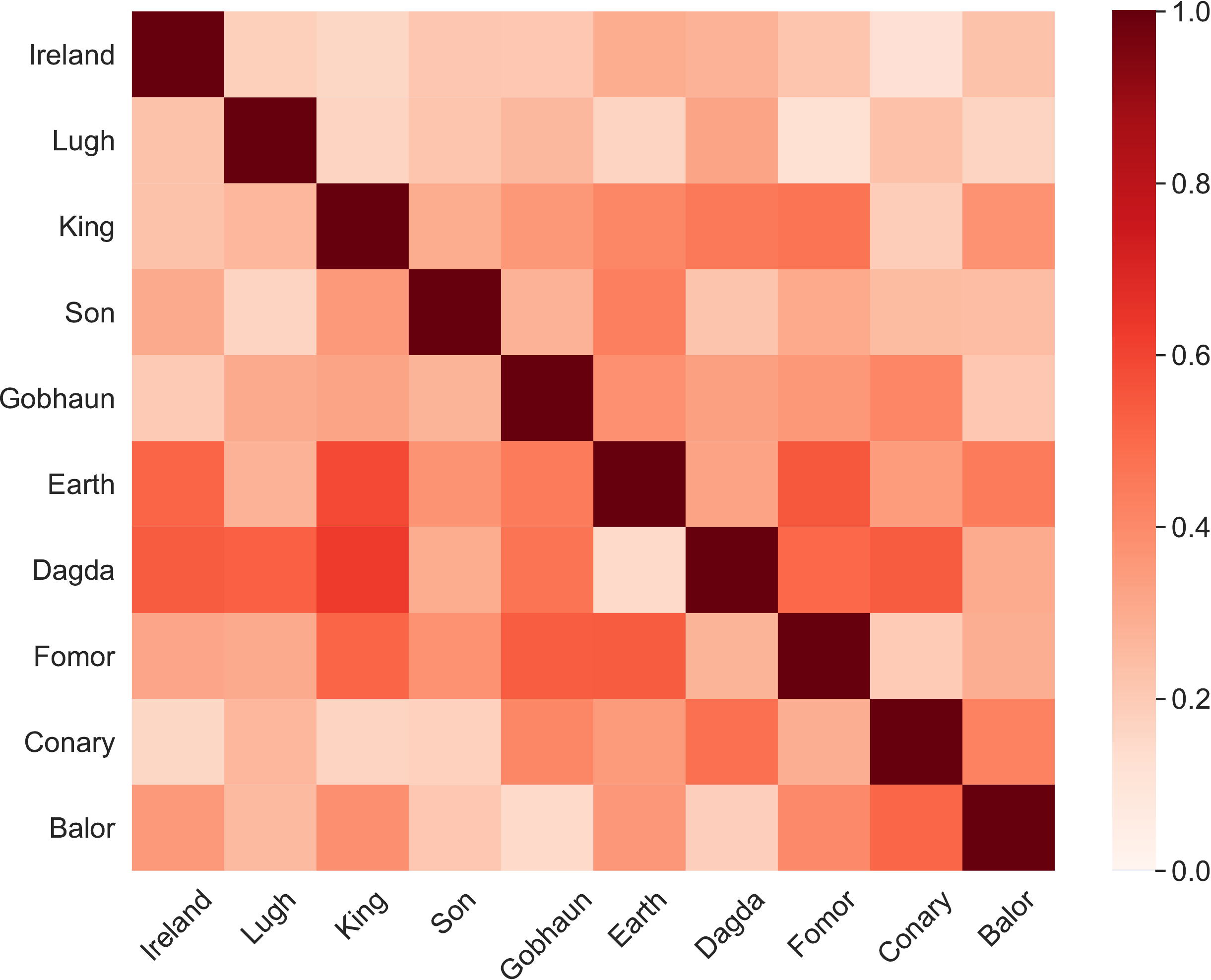}}\\
	\subfloat[]{\includegraphics[width=0.49\linewidth]{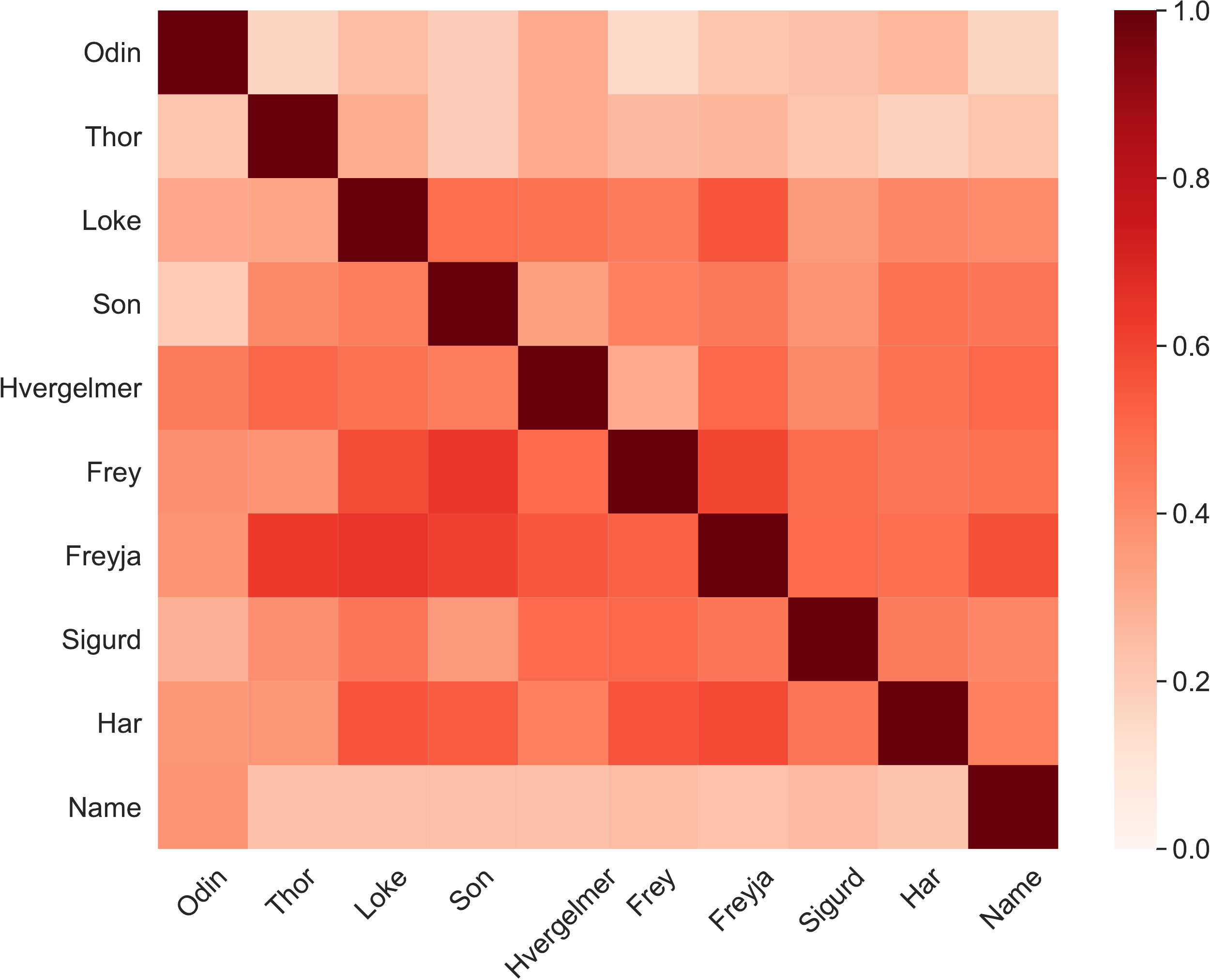}}
	\subfloat[]{\includegraphics[width=0.49\linewidth]{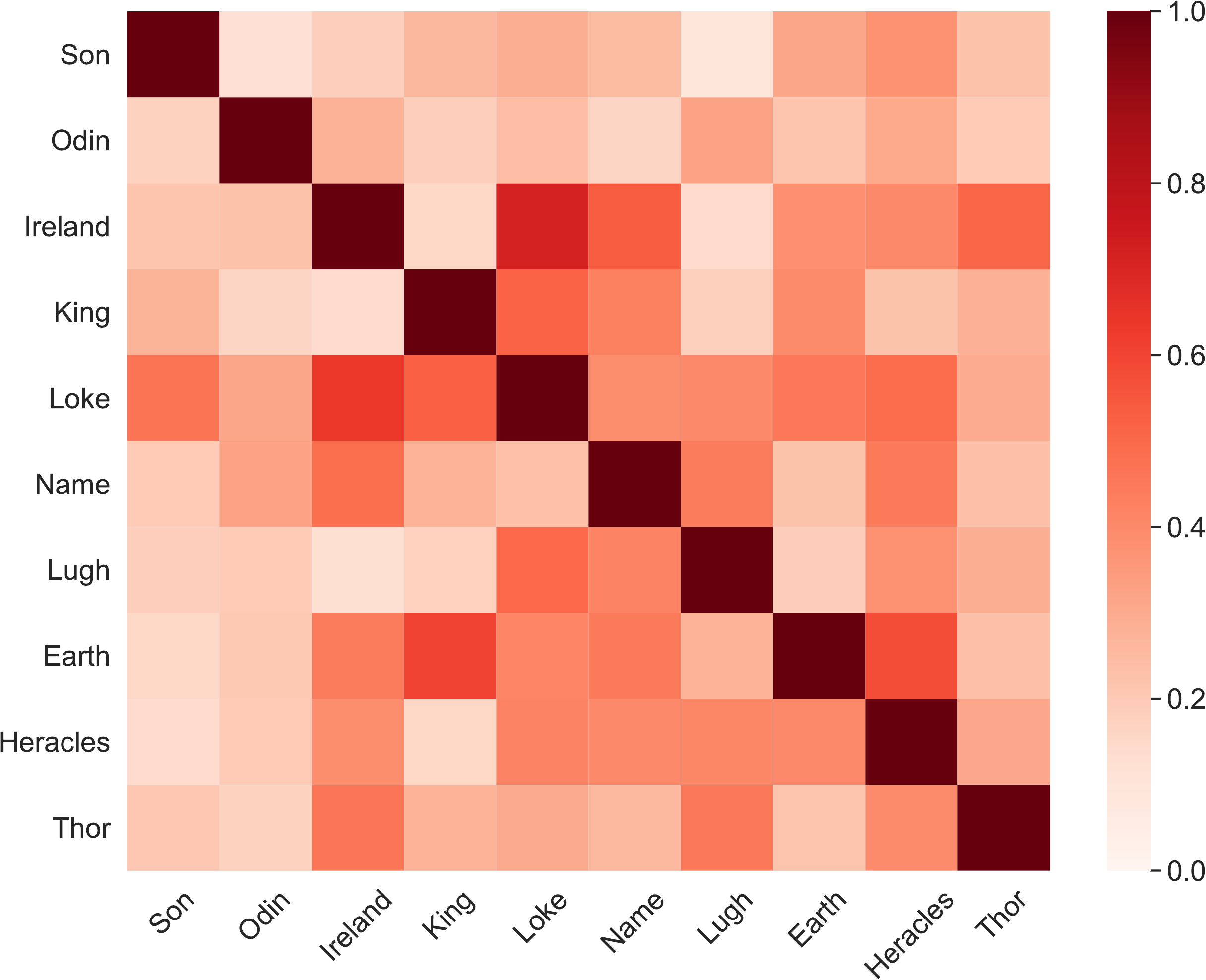}}
	\caption{\textbf{Semantic affinities in all the networks studied}. We chose the 10 most repeated entities in each text to compare themselves. \textbf{a} \emph{Greek myths} network. \textbf{b} \emph{Celtic Wonder-Tales} network. \textbf{c} \emph{Younger Edda} network. \textbf{d} Fused myths network.}
	\label{fig:distances}
\end{figure}

In Fig. \ref{fig:distances} we show how the 10 most repeated entities in each text, ordered according to semantic value, relate to each other in terms of semantic affinity and showcase some of the most interesting actors to study. There are some relationships to remark:

\begin{itemize}
	\item ``Psyche'', the impersonation of the human soul and lover of ``Eros'', receives significant semantic affinity from ``Persephone'', who is the wife of ``Hades'', and ``Jason''. There is a direct connection between ``Persephone'' and ``Psyche'', as they both appear in the same story, and both are the wife of a god and both are connected to the underworld. However, there is not straightforward connection with ``Jason''.
	\item ``Apollo'' sends the most of his affinity to ``Heracles'', who is also a character related to many different virtues, and receives the most from ``Jason''. ``Jason'' is also an important hero, so it is natural that he is connected to the god of virtue. Besides, ``Jason'''s mother was a lover of ``Apollo'', which might imply deeper connections between these two characters.
	\item ``Zeus'' is connected more strongly to human characters than other gods and its most important connection is with ``Psyche'', the personification of the human soul.
\end{itemize}

In Fig. \ref{fig:distances}(b) we show how the top 10 entities according to $S$ value in the Celtic mythology network relate to each other in terms of semantic affinity. Some of the findings in this figure are:

\begin{itemize}
	\item ``Dagda'', the sun god, emits most semantic affinity to ``King''and ``Ireland'', which suggests a relationship between earthly and divine mandates. Besides, ``Dagda'' is heavily entwined with ``Ireland'' but not with ``Earth'', which implies a negative connotation for ``Earth''. This might be in line with the idea that good things are ``heavenly'' things and ``bad'' things are more ``earthly''.
	\item ``Earth'' emits a lot of semantic value to ``King'', reinforcing again the bond between the earth and the ruler.
	\item  ``Balor'', the king of the Fomorians, emits most of his affinity to ``Conary M\'{o}r'', which is a prototype for a good king. However, this king also breaks three sacred oaths in his story and this might connect him with negative characters such as ``Balor''. ``Balor'' also emits significant semantic value to its tribe, ``Fomor'', which is expected, but also to ``Earth'' and ``Ireland''.
\end{itemize} 

In Fig. \ref{fig:distances}(c) we show how the top 10 entities in $S$ value in the Nordic mythology network relate to each other in terms of semantic affinity and remark some of the most interesting actors to study. We can observe that:

\begin{itemize}
	\item Both ``Odin'' and ``Thor'' have very high semantic affinity values compared to other actors. This is probably due to the fact that these gods have many attributions. In the case of ``Name'', which does not have neither high semantic affinity values, it does have a higher semantic affinity to ``Odin'', probably for the high number of different names that this god has in the texts.
	\item ``Freyja'' emits and receives significant affinity from ``Loki''. ``Loki'' is the responsible for the death of the almost invincible god ``Baldr'', who is also ``Freyja'''s son. ``Freyja'' is considered the leader of the Valkyries and takes half of the fallen to her own afterlife field. This high affinity value here might indicate that the death theme is in fact a very important bond between them. 
	\item ``Frey'', one of the most important Vanir gods, sends the most affinity to the actor ``Son''. ``Frey'' is usually associated with sacral kingship and his name derived phonetically from old Norse means ``Lord''. This probably says that the original sacramental attributions of this Vanir god were abandoned in favor of the attributes of the more popular \AE sir.
	\item ``Hvergelmer'' is the fountain in Nifelheim, the reign of the dead from which all rivers are born, and it is mostly associated with chaos. ``Hvergelmer'' sends and receives a significant amount of affinity from ``Freyja'' and ``Loki'', and both of them showed certain relationship with death. Besides, just as in the case of ``Heracles'' and ``Hydra'', the most emitted semantic affinity is to ``Thor'', which is considered to be associated to order.
\end{itemize}

In Fig. \ref{fig:distances}(d) we did the analogous experiments for the fusion network of the three mythologies. Among other possible interesting relationships, it is notable that:

\begin{itemize}
	\item The highest affinity of ``Lugh'' is ``Loki''. This is a remarkable result, as there have been many studies discussing a possible a relationship between these two gods \cite{ewing1995birth}.
	\item A strong semantic affinity between ``Ireland'' and ``Loki'', in both directions, and between ``Ireland'' and ``Thor'', to a lesser extent. This might be due to ``Ireland'' being notably close to ``Lugh'', who is a god closely entwined to both ``Odin'' and ``Loki'', and because all of them are symbols related to authority in their original stories.
	\item ``Earth'' is notably affine to ``King'', which means that the strong tie between the land and the ruler present in the \emph{Celtic Wonder-Tales} network is also present in the other two.
	
\end{itemize}

\subsection{Semantic affinity compared to other affinities in relevant actors}

\begin{figure}
	\centering
	\adjustbox{max width=\linewidth}{
	\begin{tabular}{ccc}
		\multicolumn{3}{c}{\textbf{Greek Myths}}   \\
		\midrule
		\multicolumn{3}{c}{Zeus}   \\
		\midrule
		\includegraphics[width=0.5\linewidth]{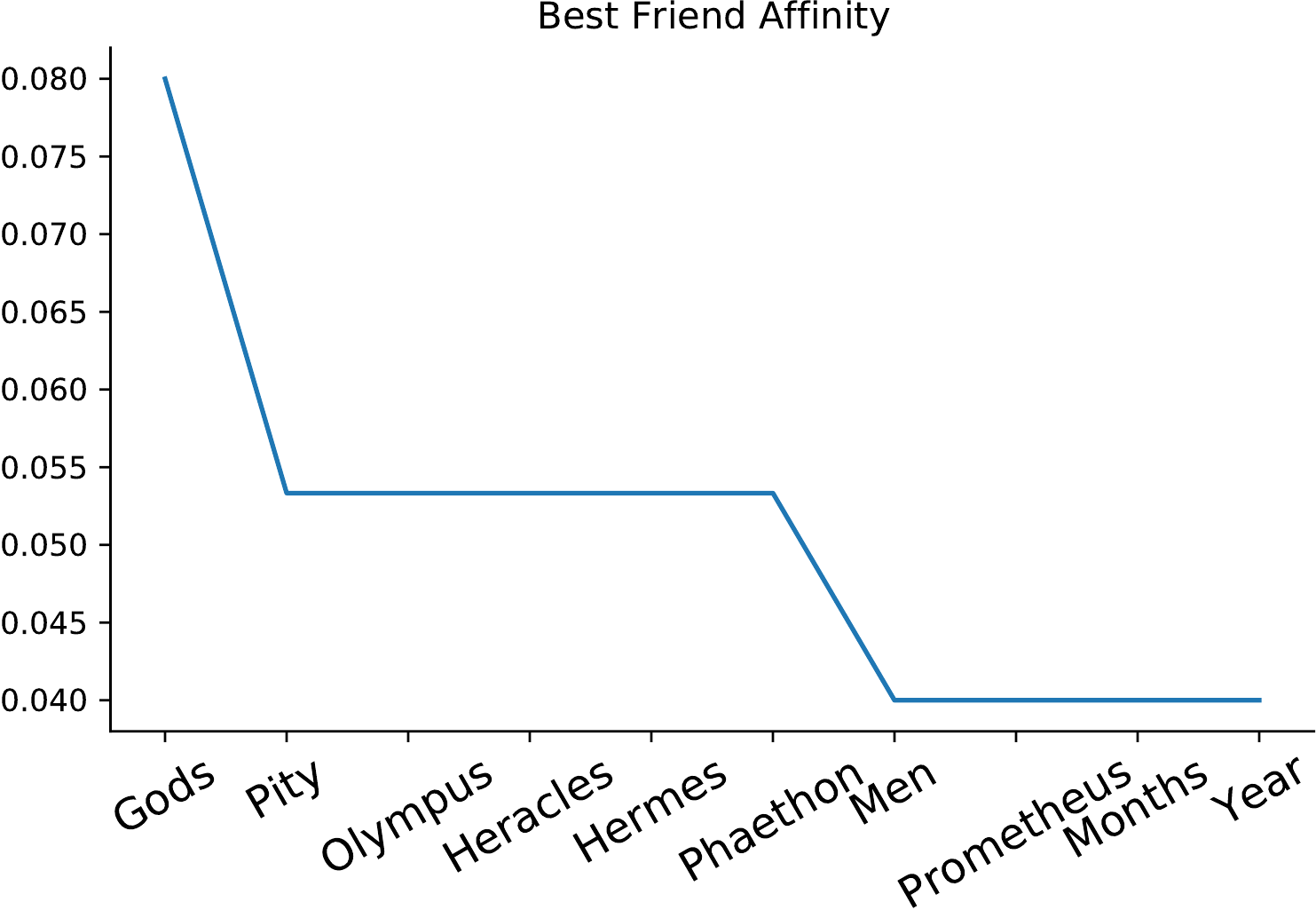} & \includegraphics[width=0.5\linewidth]{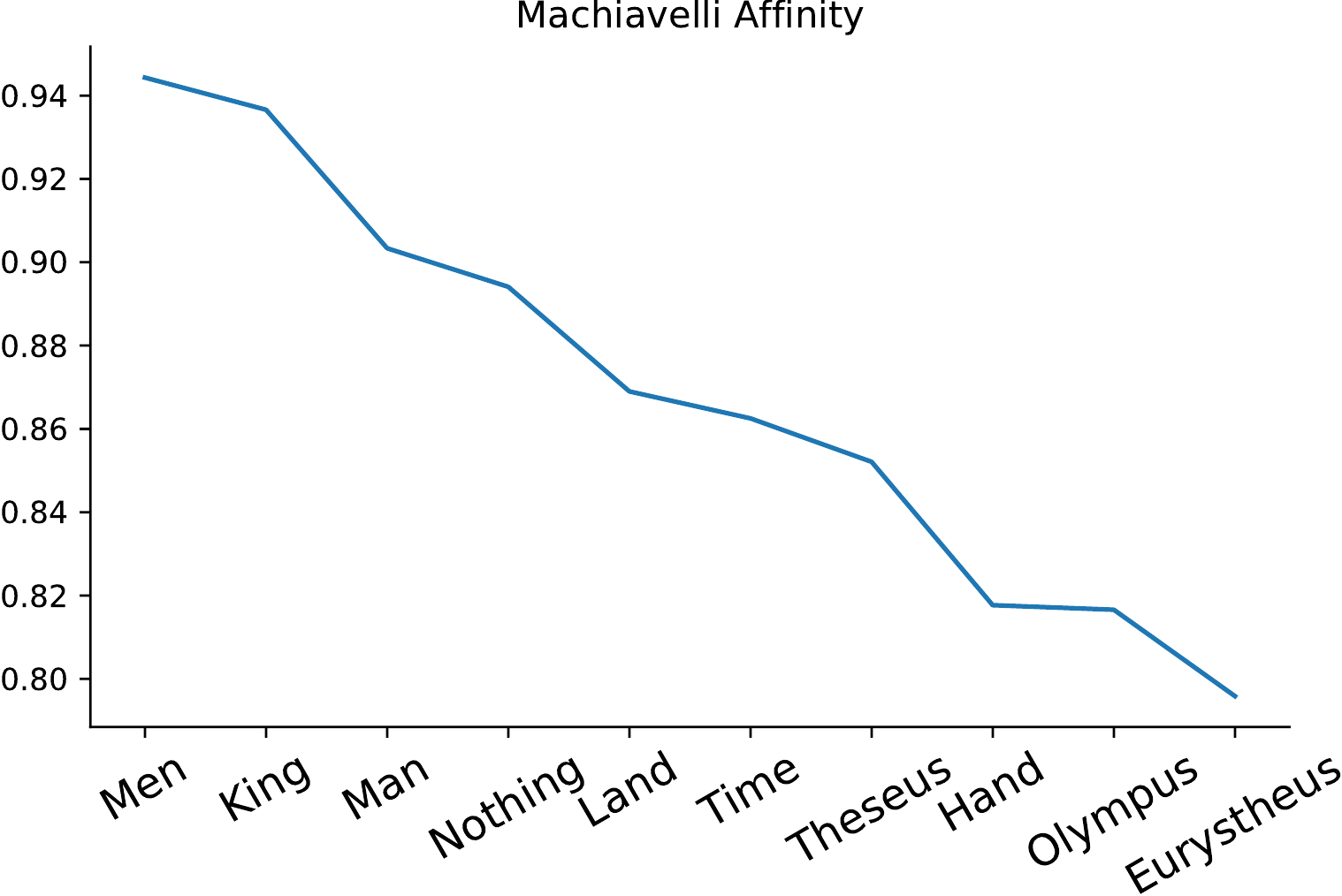} & 
		\includegraphics[width=0.5\linewidth]{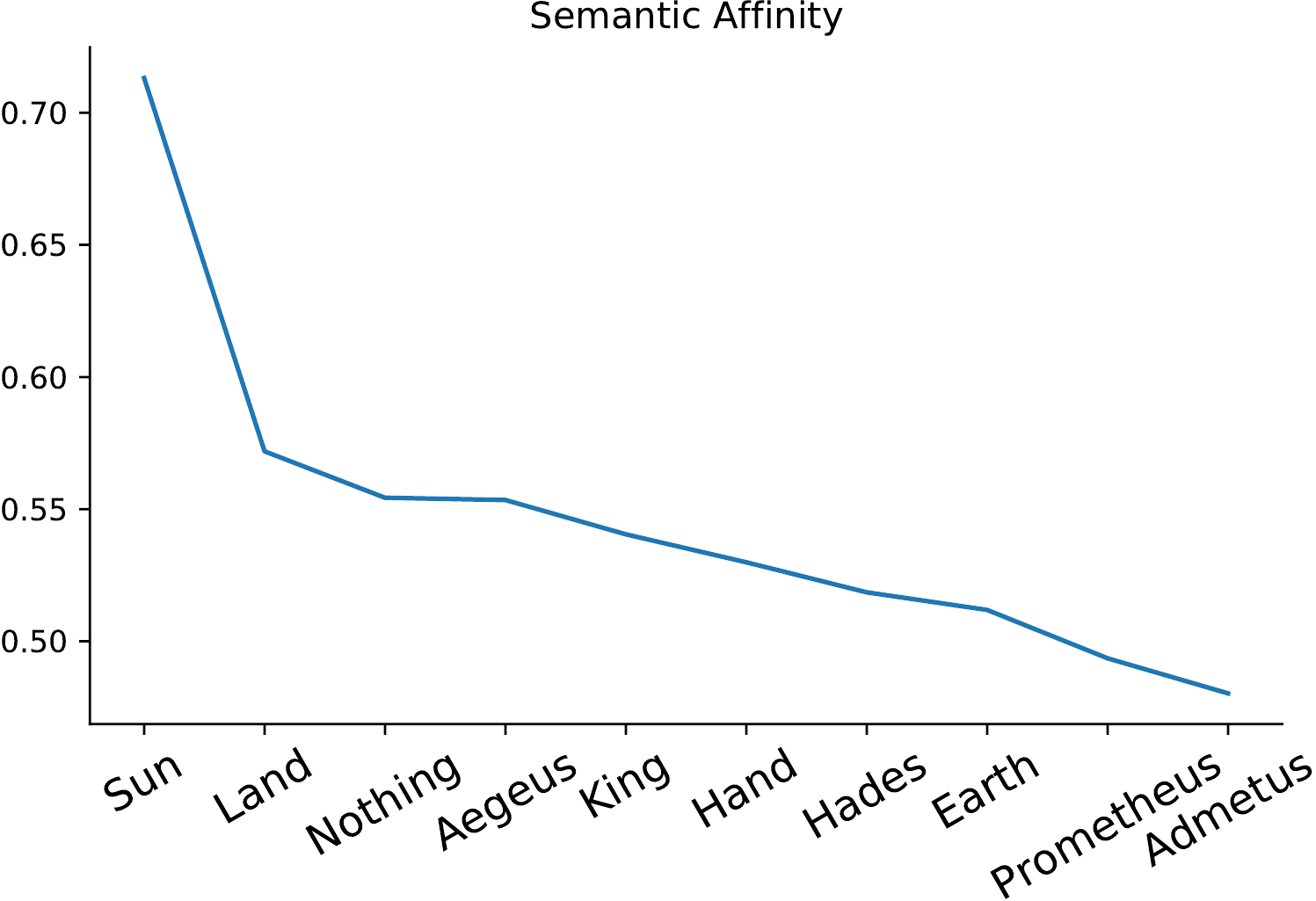} \\
		\midrule
		\multicolumn{3}{c}{Athene}   \\
		\midrule
		\includegraphics[width=0.5\linewidth]{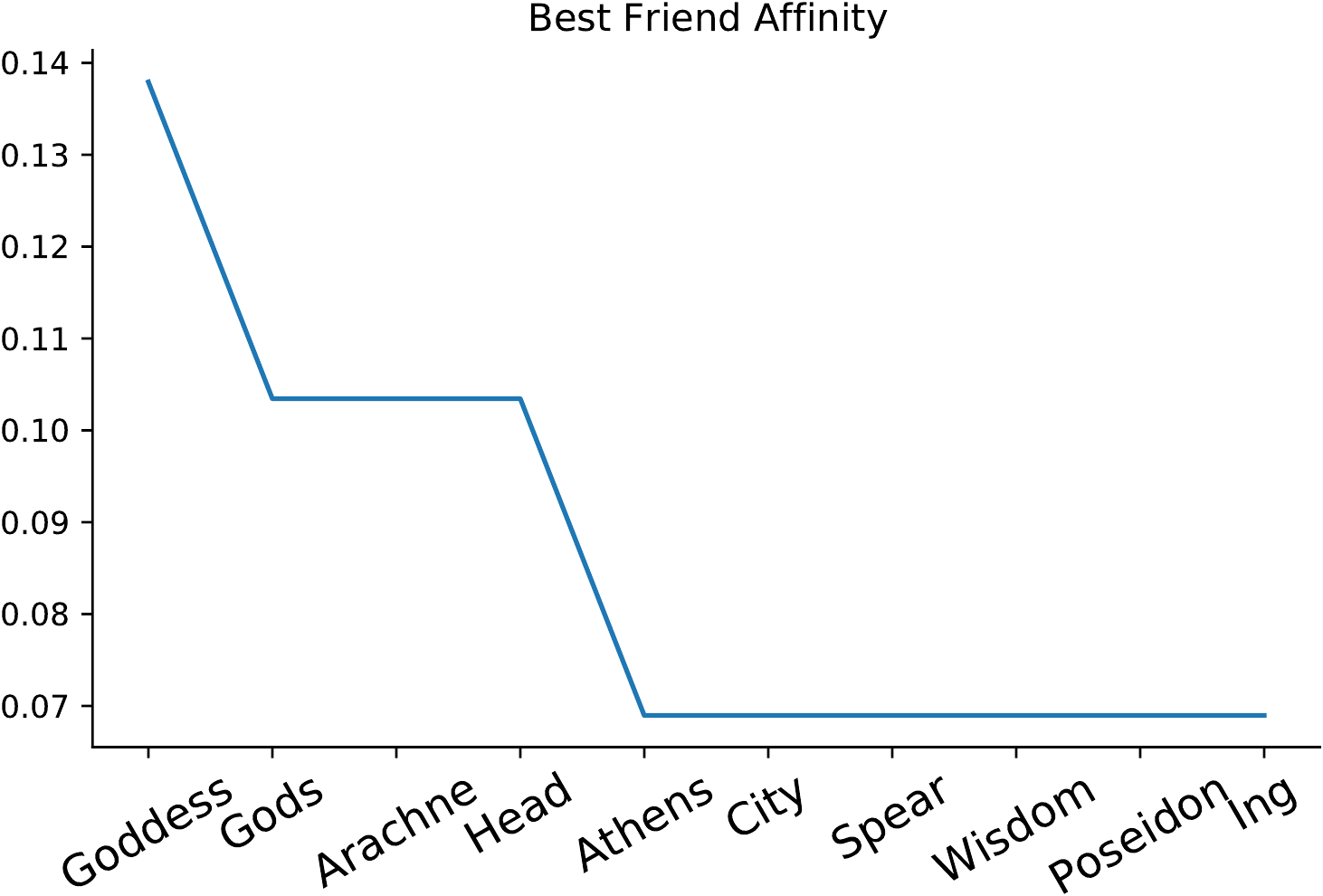} & \includegraphics[width=0.5\linewidth]{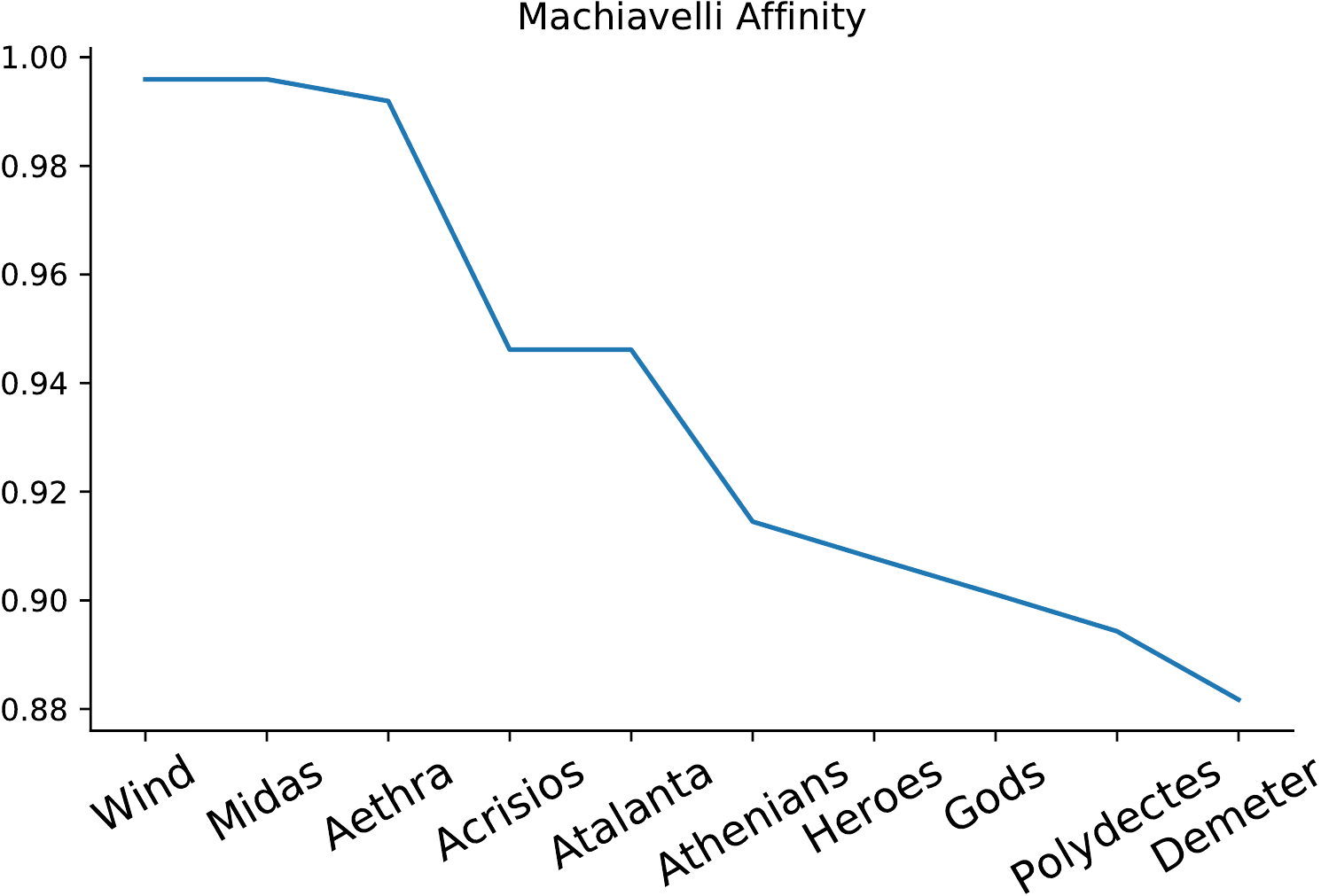} & 
		\includegraphics[width=0.5\linewidth]{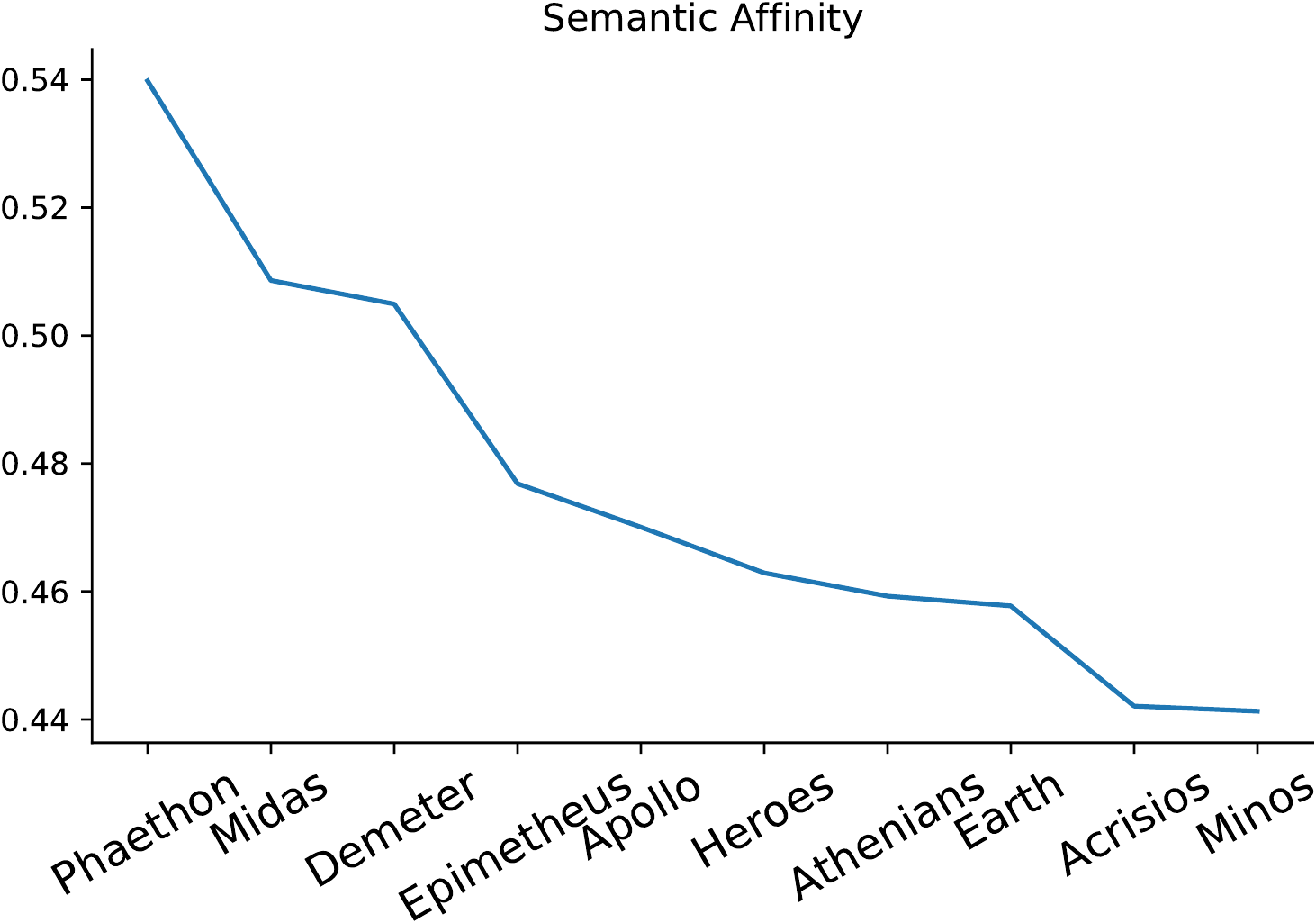} \\
		\midrule
		\multicolumn{3}{c}{Heracles}   \\
		\midrule
		\includegraphics[width=0.5\linewidth]{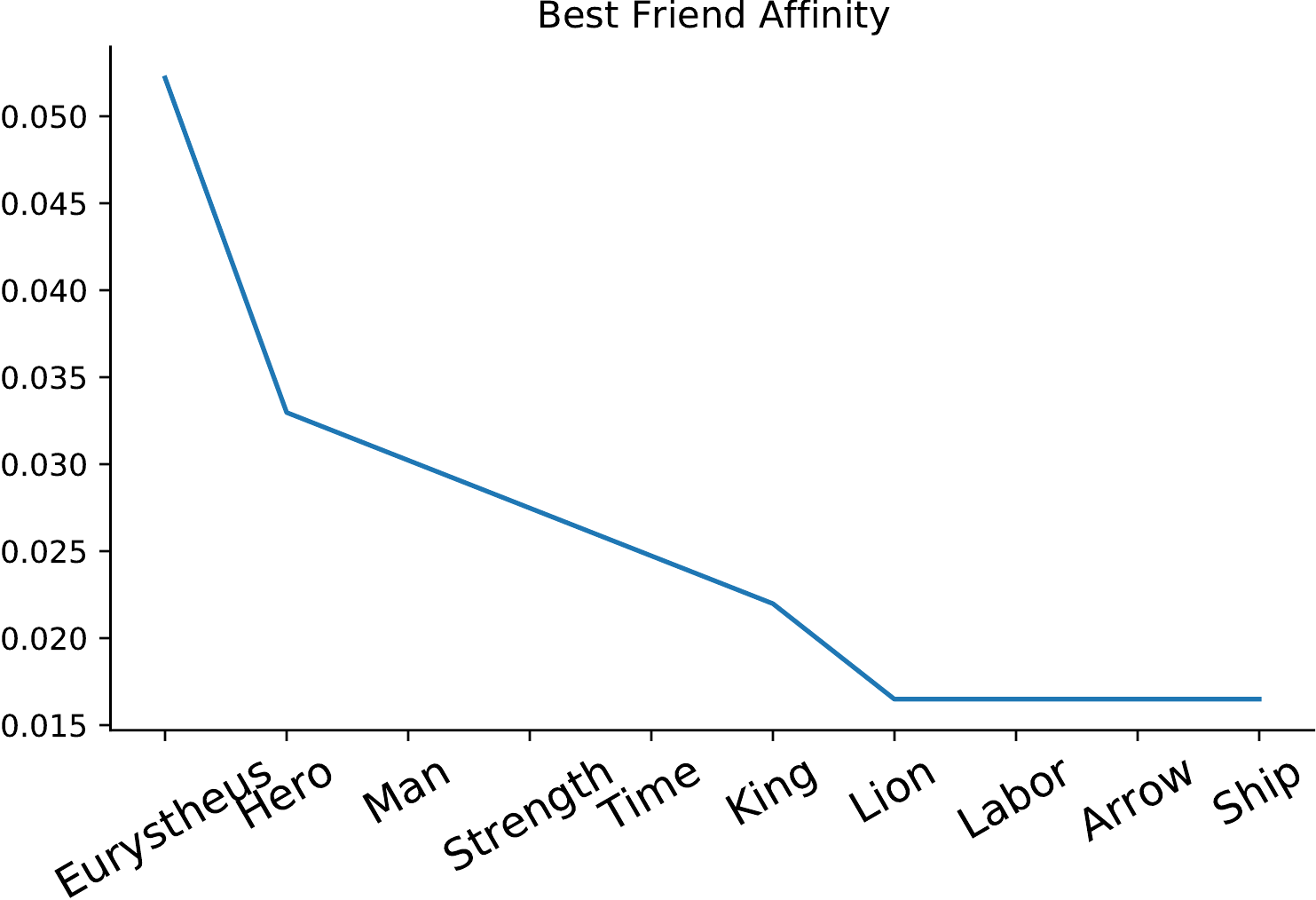} & \includegraphics[width=0.5\linewidth]{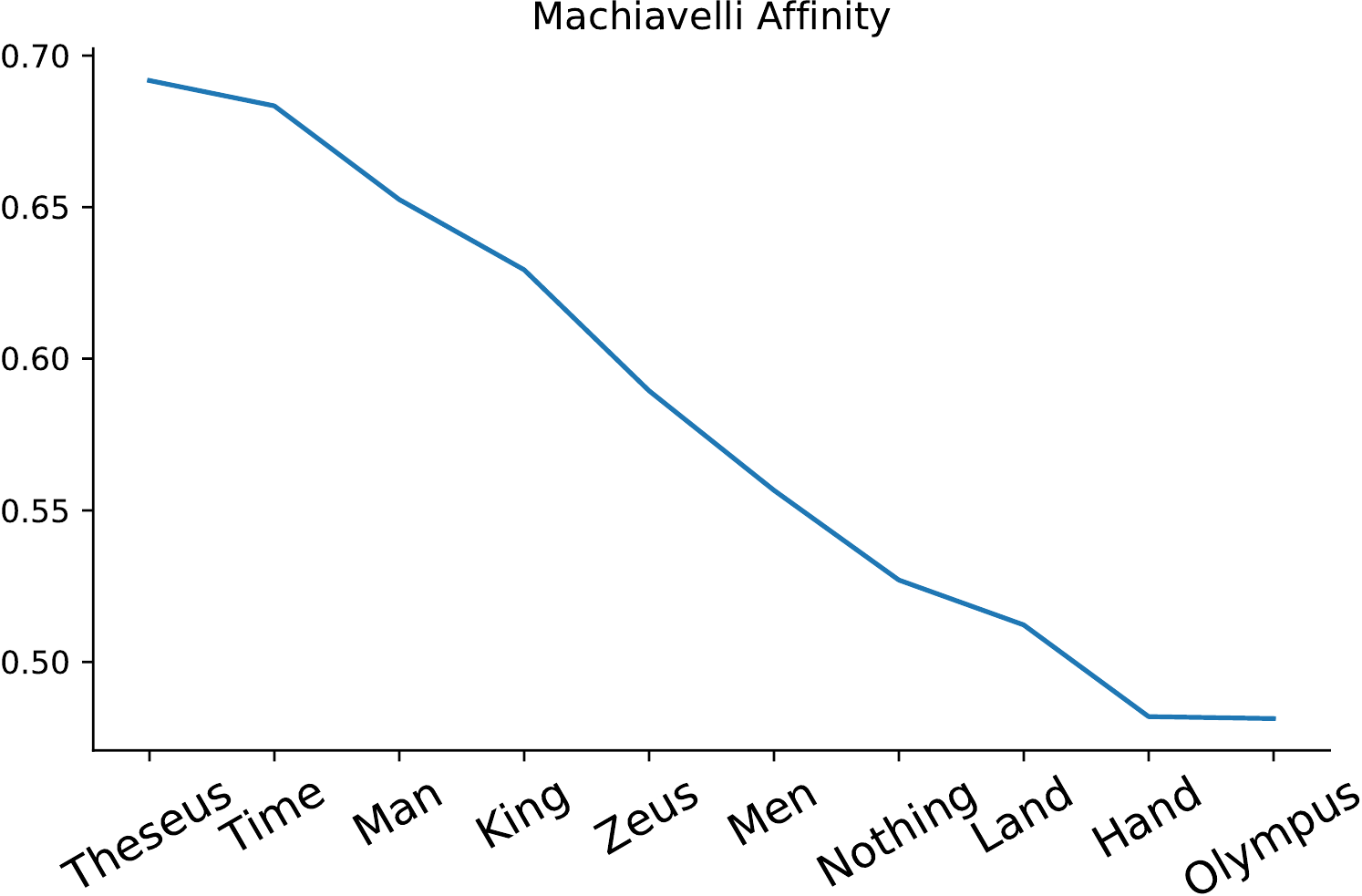} & 
		\includegraphics[width=0.5\linewidth]{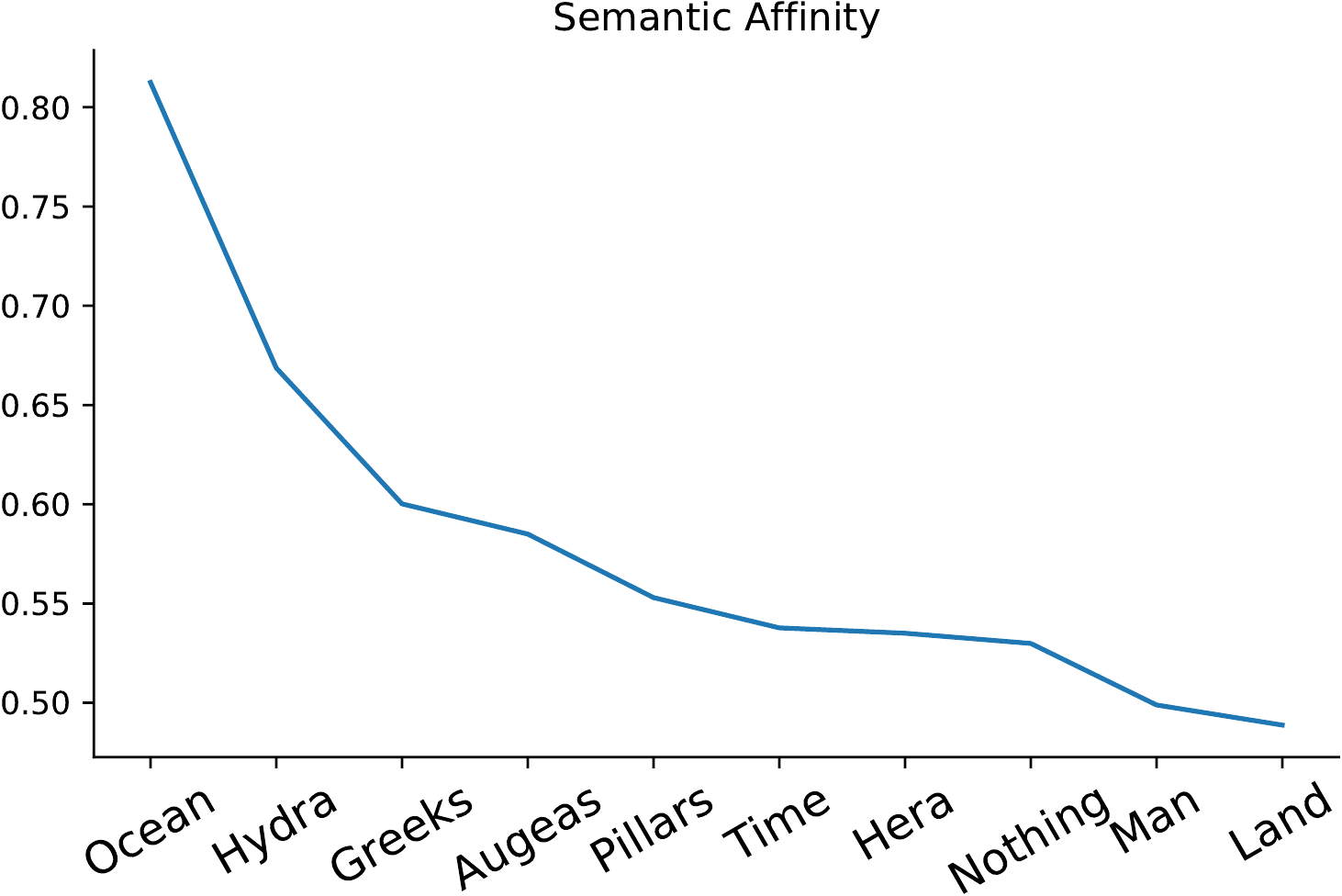} \\
		\bottomrule
	\end{tabular}
}
	\caption{\textbf{Study of affinities for three key characters in \emph{Greek myths}.} Top 10 affinity values for the best friend, Machiavelli, and semantic affinity for ``Zeus'', ``Athene'', and ``Heracles'' in the \emph{Greek Myths} network.}
	\label{fig:zeus}
\end{figure}

\begin{figure}
	\centering
	\adjustbox{max width=\linewidth}{
	\begin{tabular}{ccc}
		\multicolumn{3}{c}{\textbf{Celtic Wonder-Tales}}   \\
		\midrule
		\multicolumn{3}{c}{Tuatha D\'{e} Danann}   \\
		\midrule
		\includegraphics[width=0.5\linewidth]{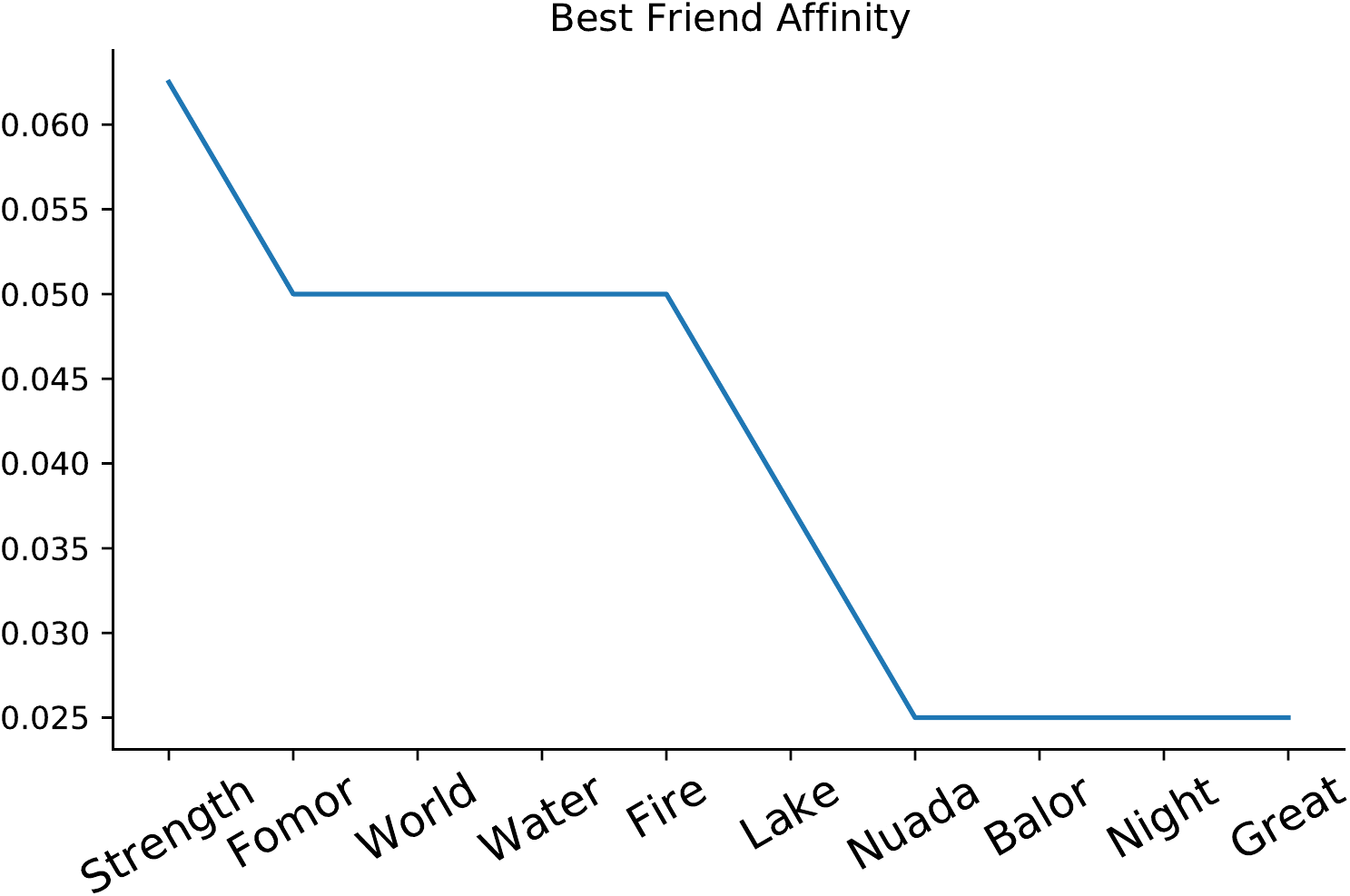} & \includegraphics[width=0.5\linewidth]{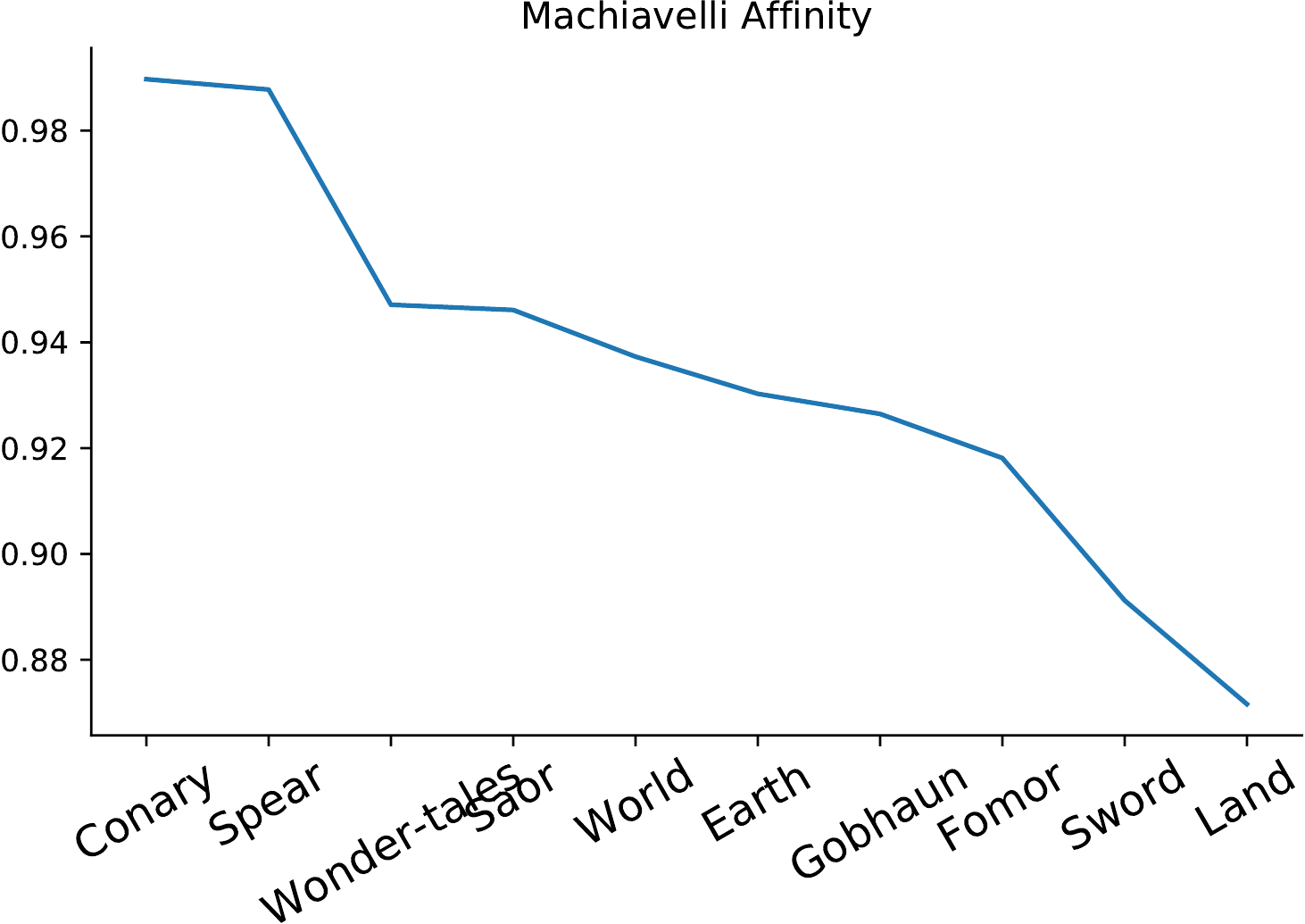} & 
		\includegraphics[width=0.5\linewidth]{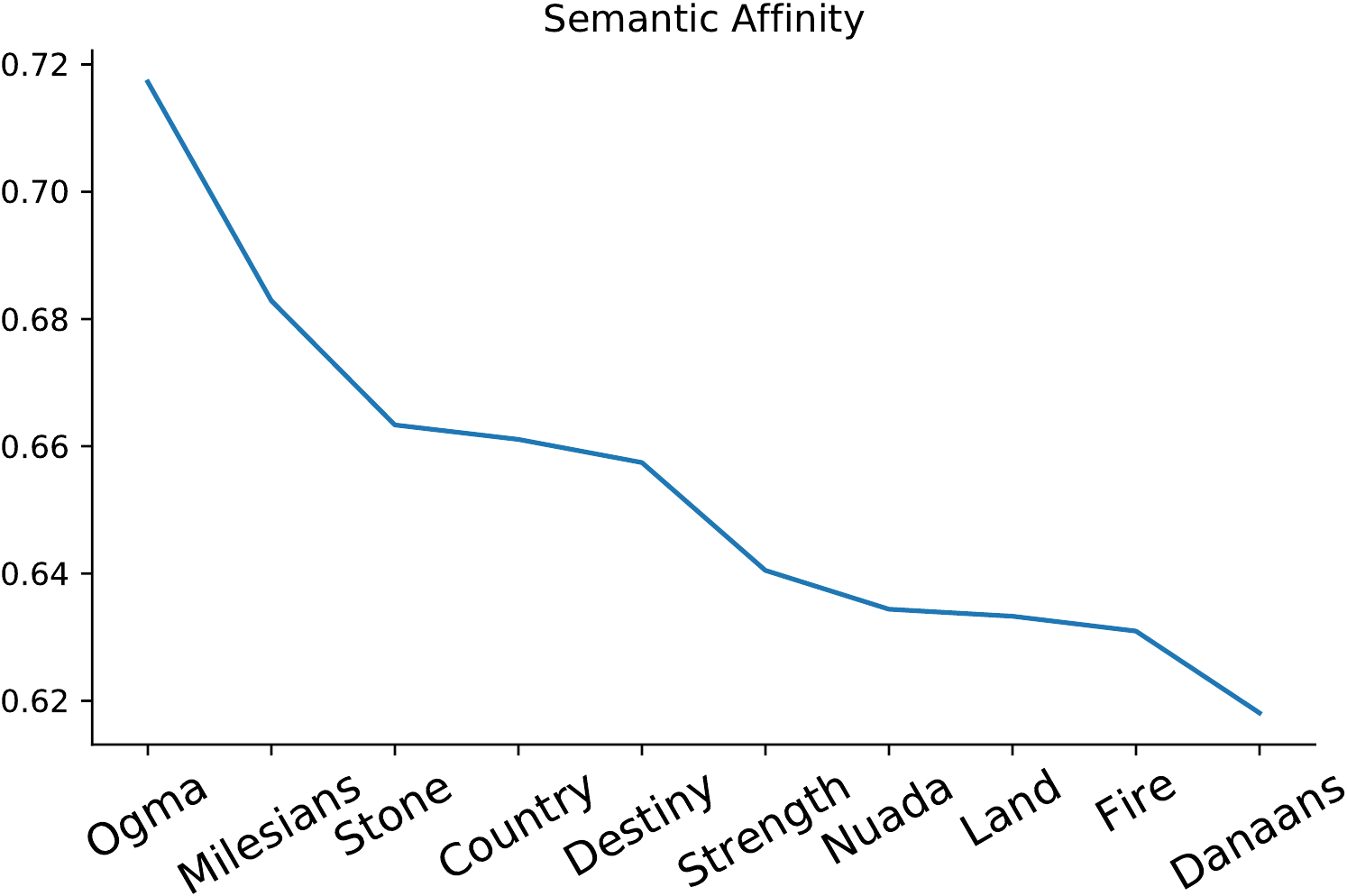} \\
		\midrule
		\multicolumn{3}{c}{Ireland}   \\
		\midrule
		\includegraphics[width=0.5\linewidth]{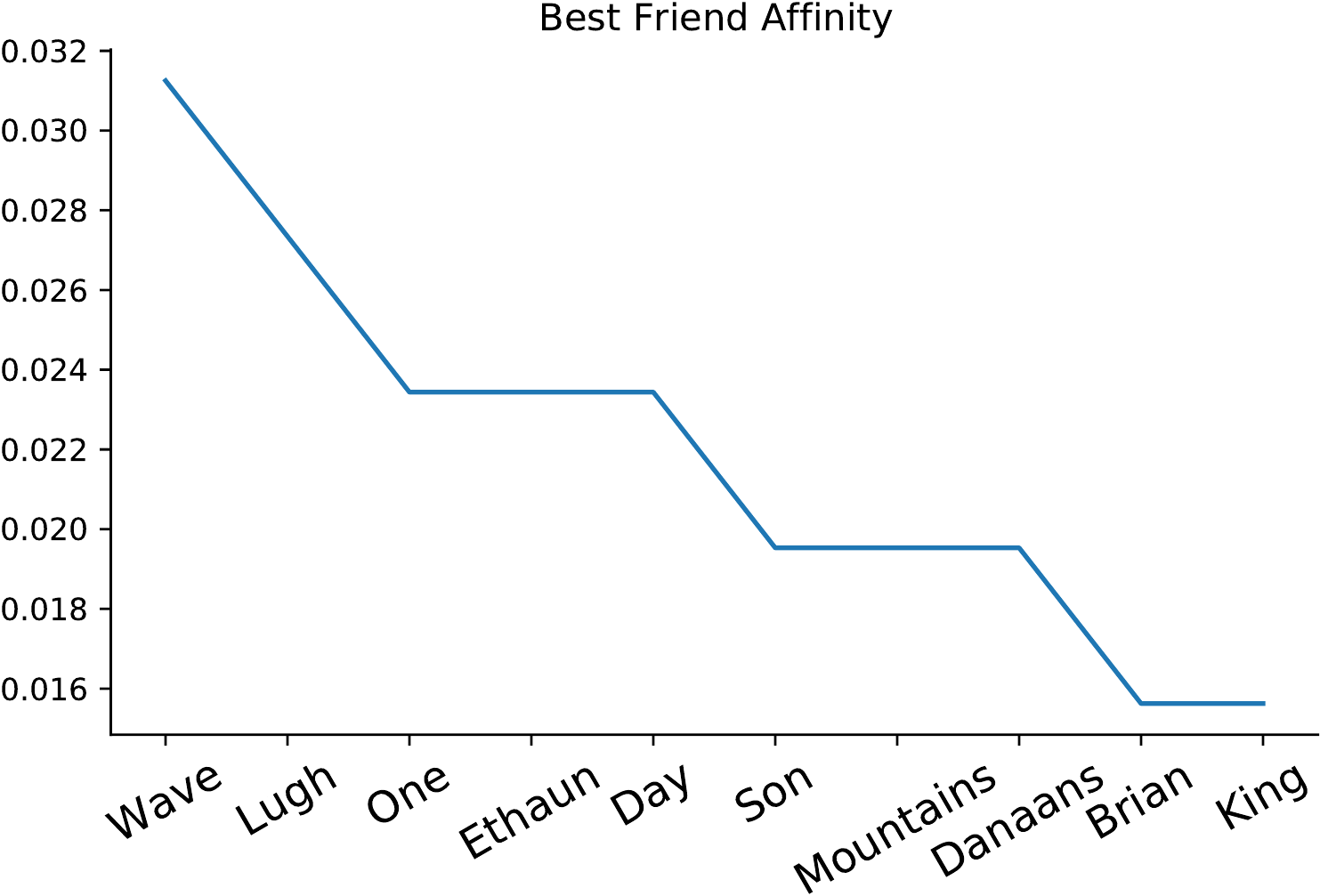} & \includegraphics[width=0.5\linewidth]{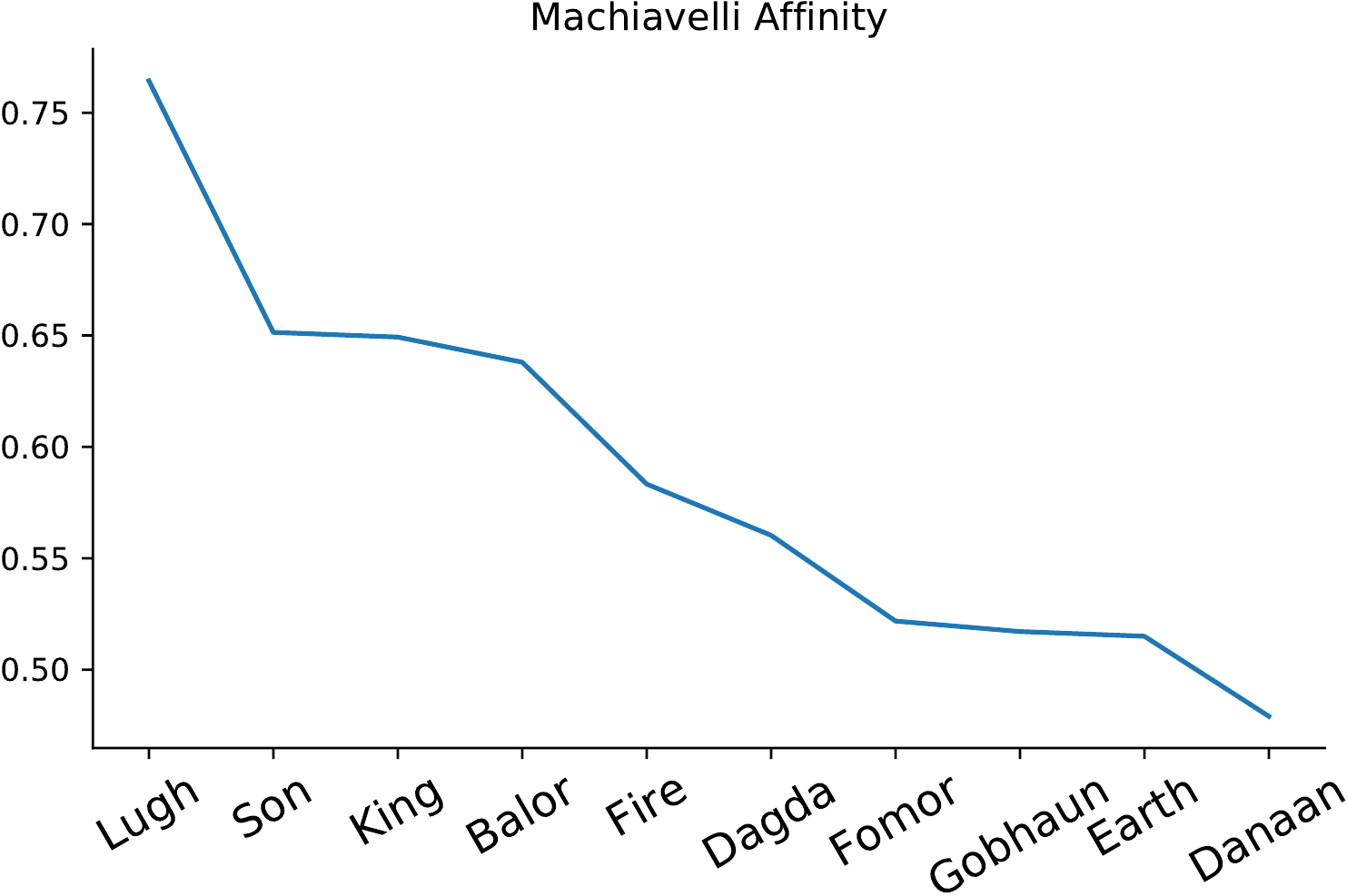} & 
		\includegraphics[width=0.5\linewidth]{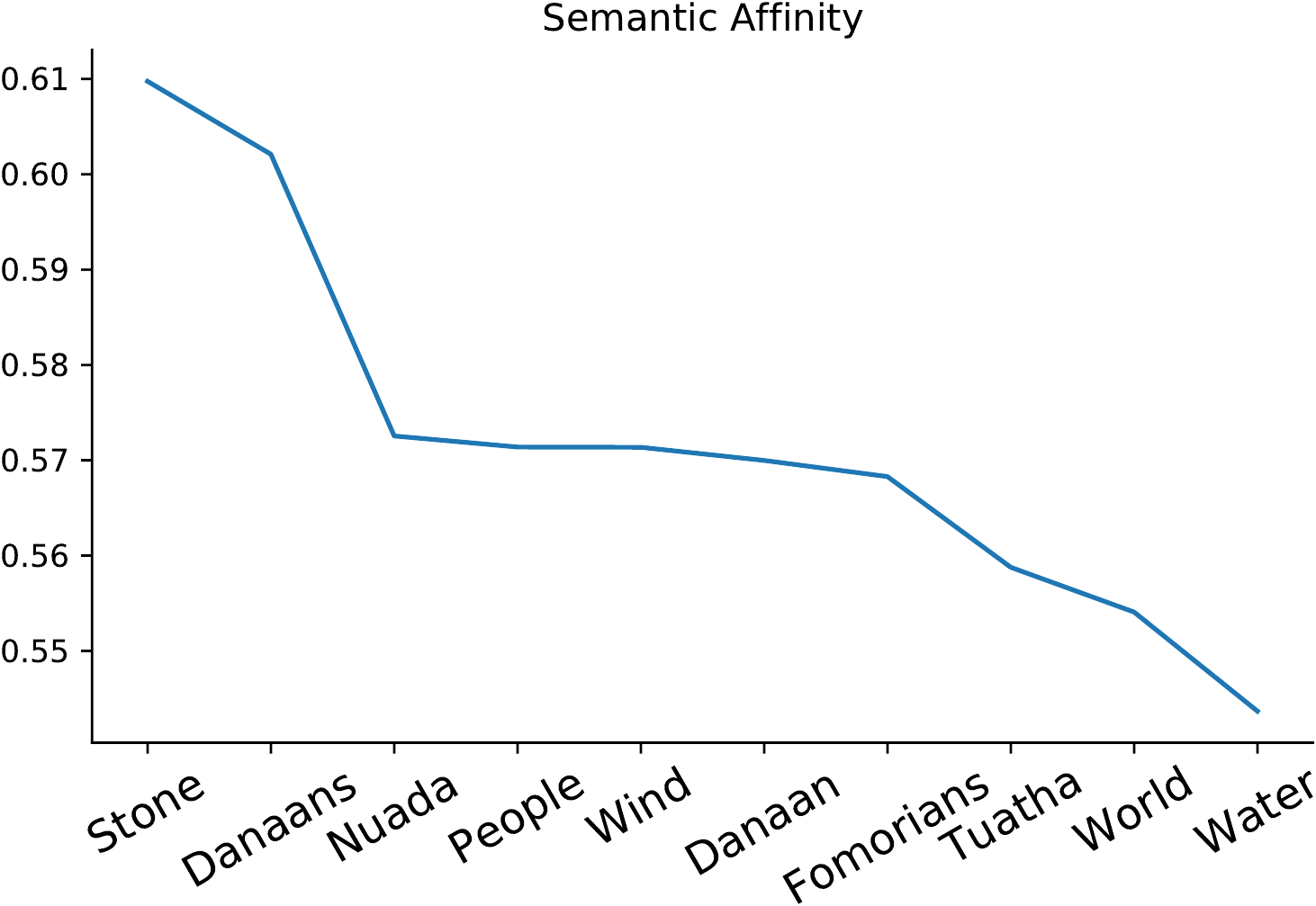} \\
		\midrule
		\multicolumn{3}{c}{Lugh}   \\
		\midrule
		\includegraphics[width=0.5\linewidth]{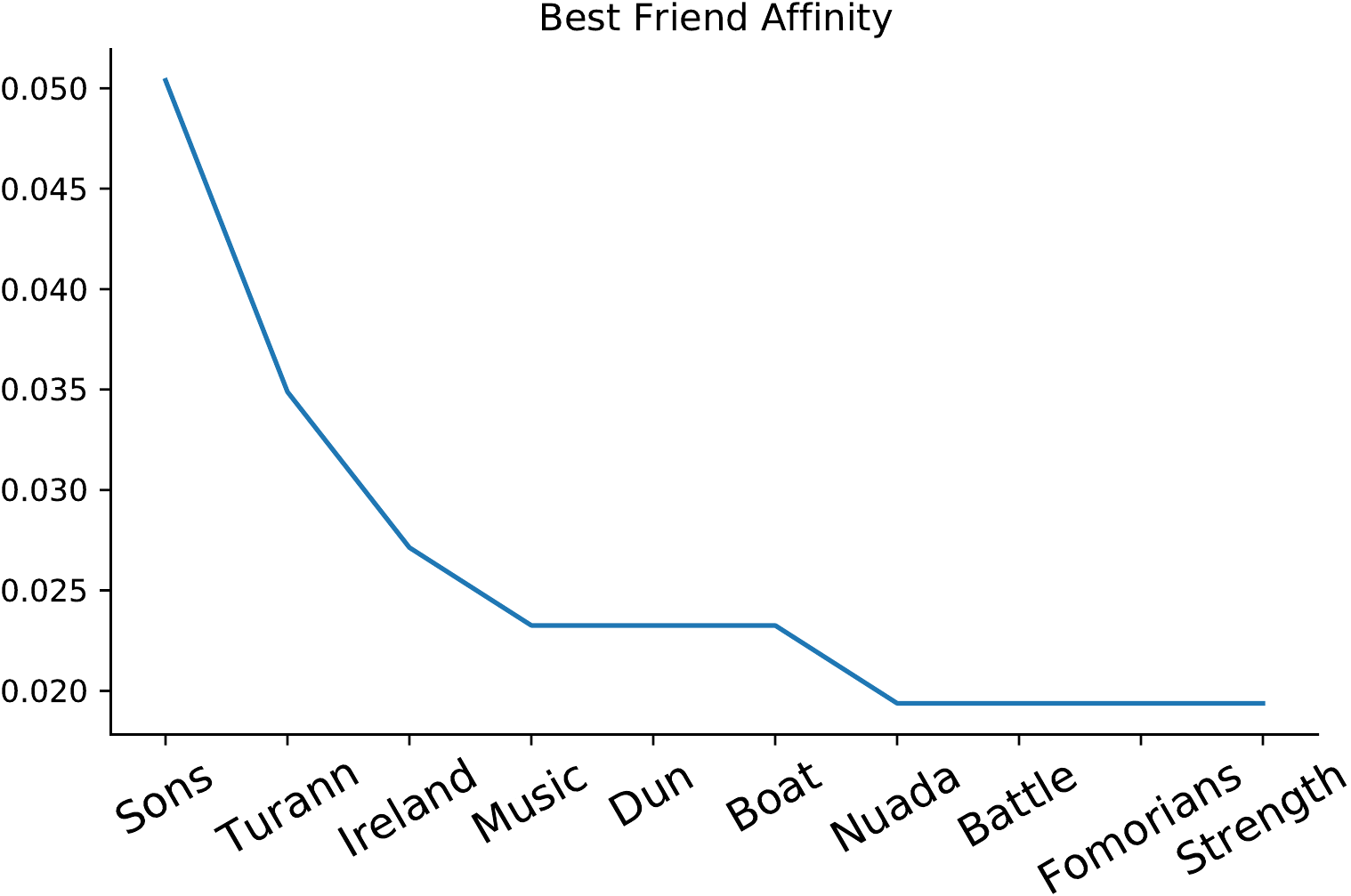} & \includegraphics[width=0.5\linewidth]{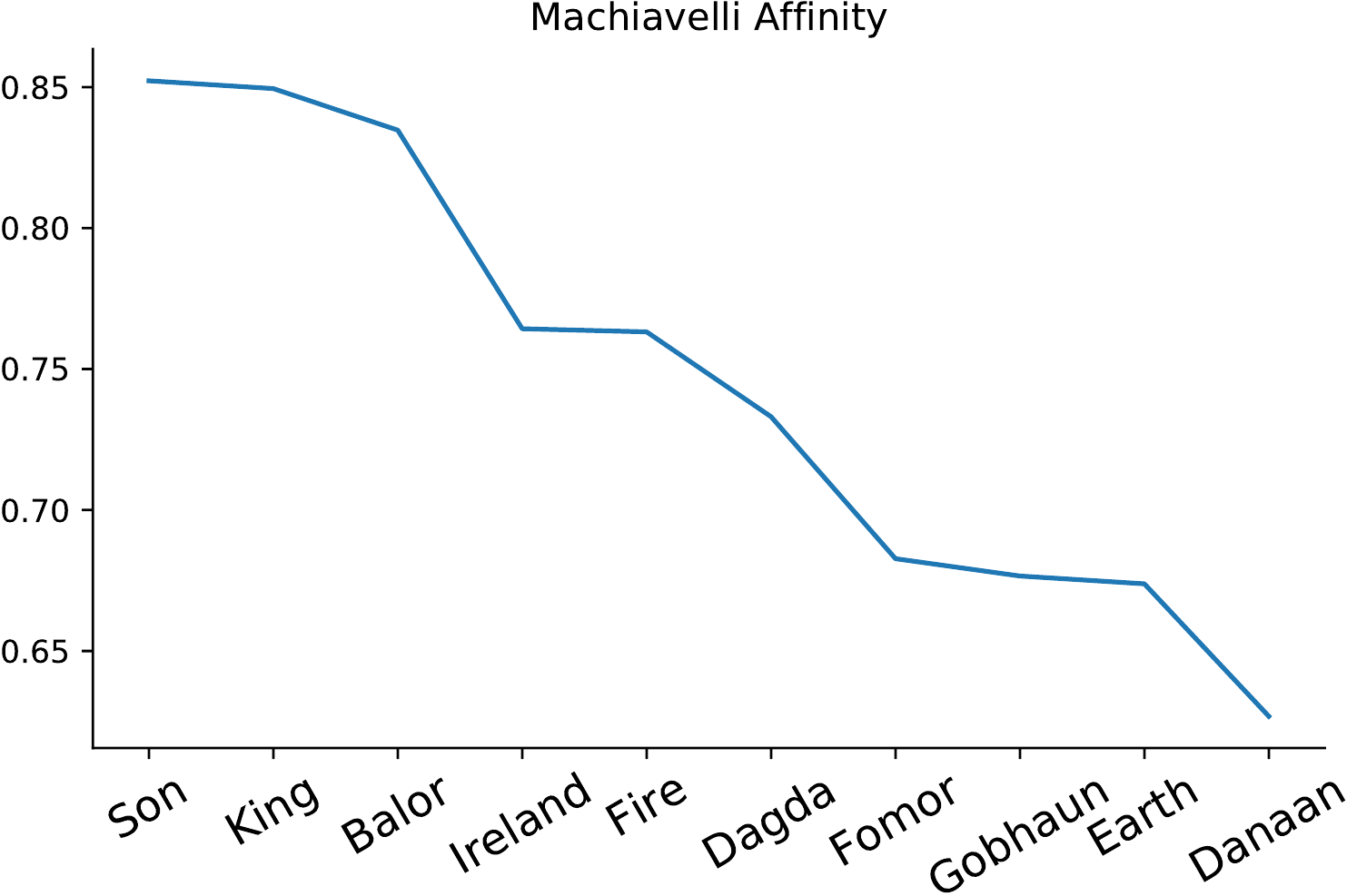} & 
		\includegraphics[width=0.5\linewidth]{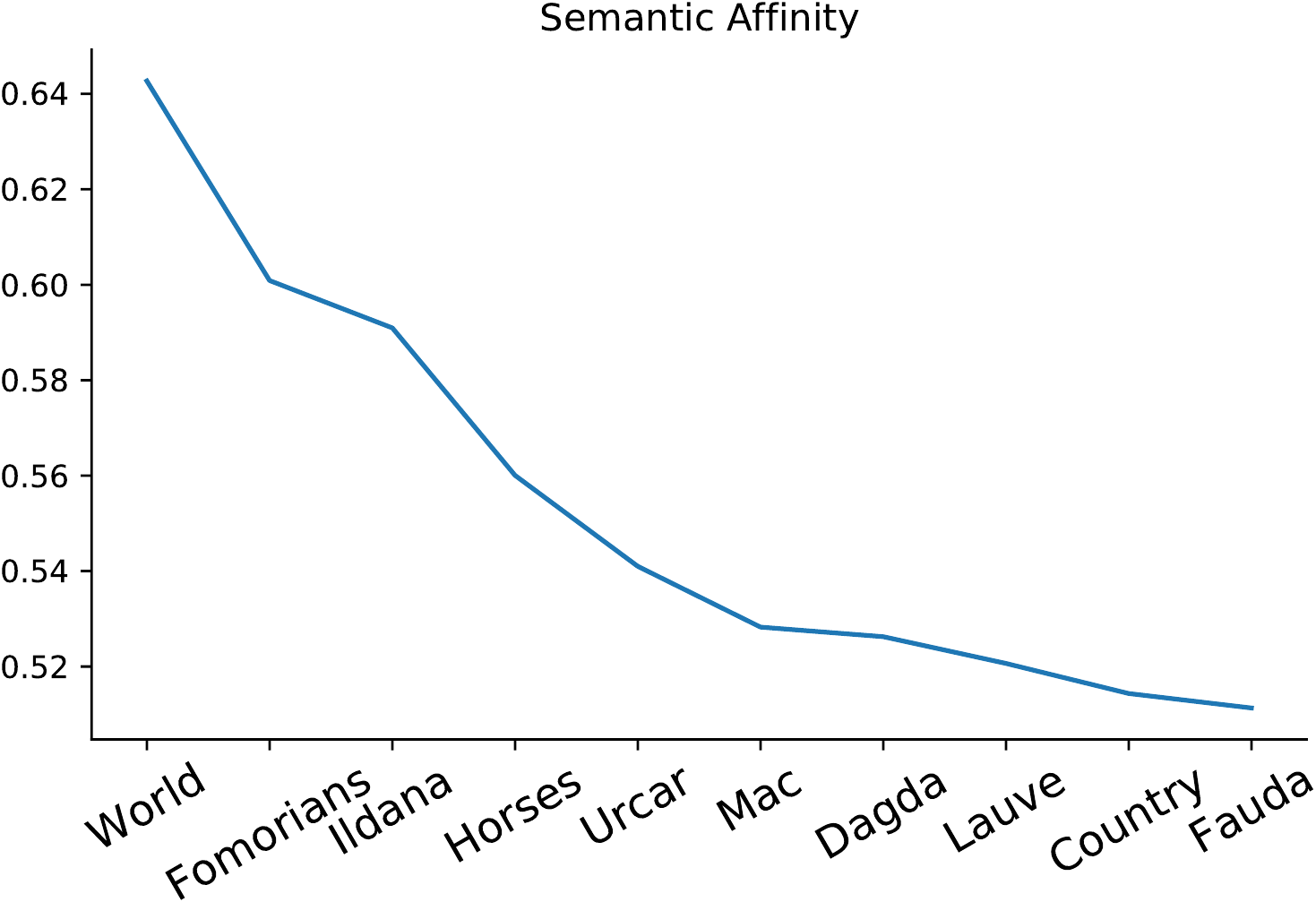} \\
		\bottomrule
	\end{tabular}
}
	\caption{\textbf{Study of affinities for three key characters in \emph{Celtic Wonder-Tales}.}  Top 10 affinity values for the best friend, Machiavelli, and semantic affinity for ``Tuatha D\'{e} Danann'', ``Ireland'', and ``Lugh''  in the \emph{Celtic Wonder-Tales} network.}
	\label{fig:danaan}
\end{figure}

\begin{figure}
	\centering
	\adjustbox{max width=\linewidth}{
	\begin{tabular}{ccc}
		\multicolumn{3}{c}{\textbf{Younger Edda}}   \\
		\midrule
		\multicolumn{3}{c}{Odin}   \\
		\midrule
		\includegraphics[width=0.5\linewidth]{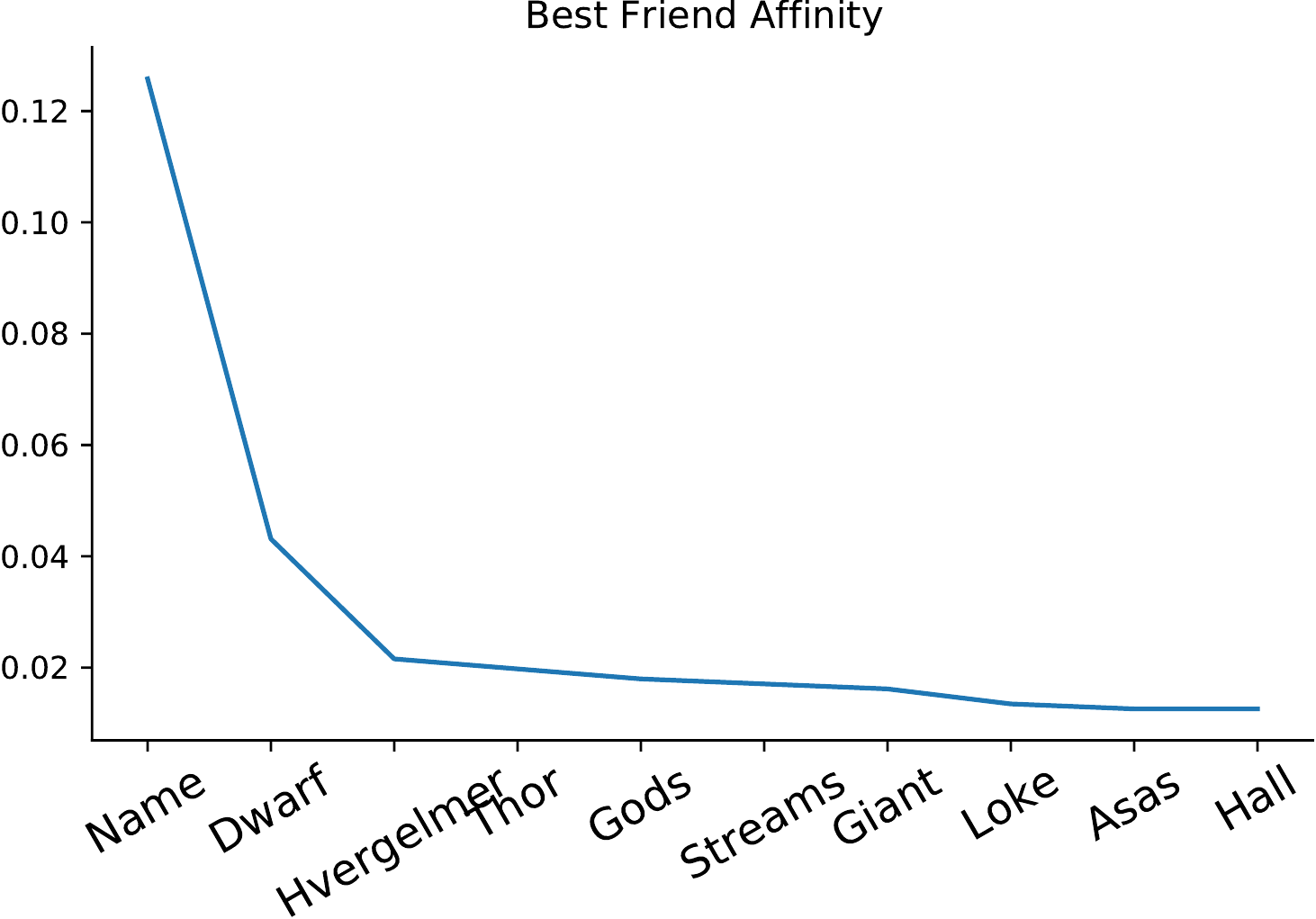} & \includegraphics[width=0.5\linewidth]{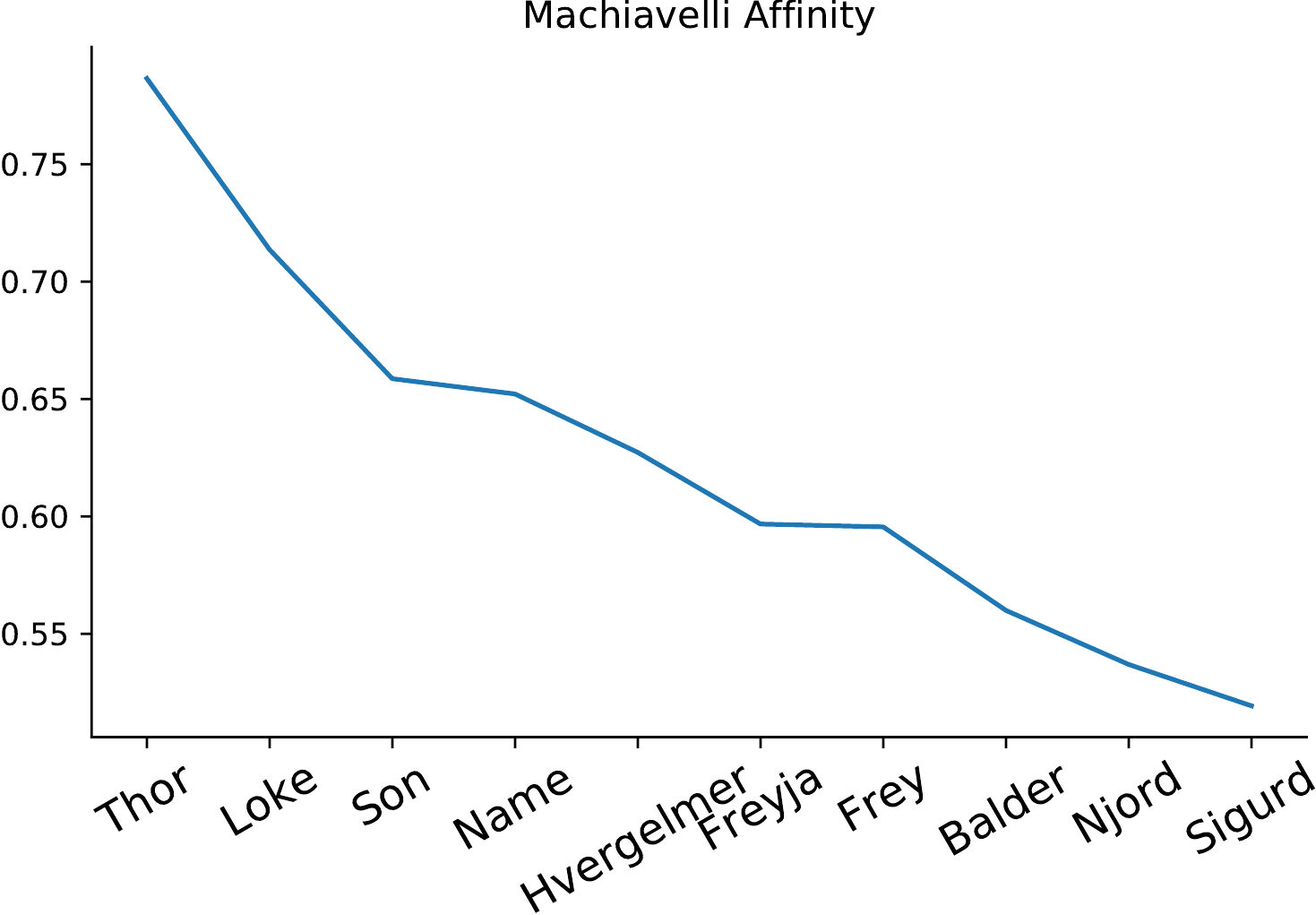} & 
		\includegraphics[width=0.5\linewidth]{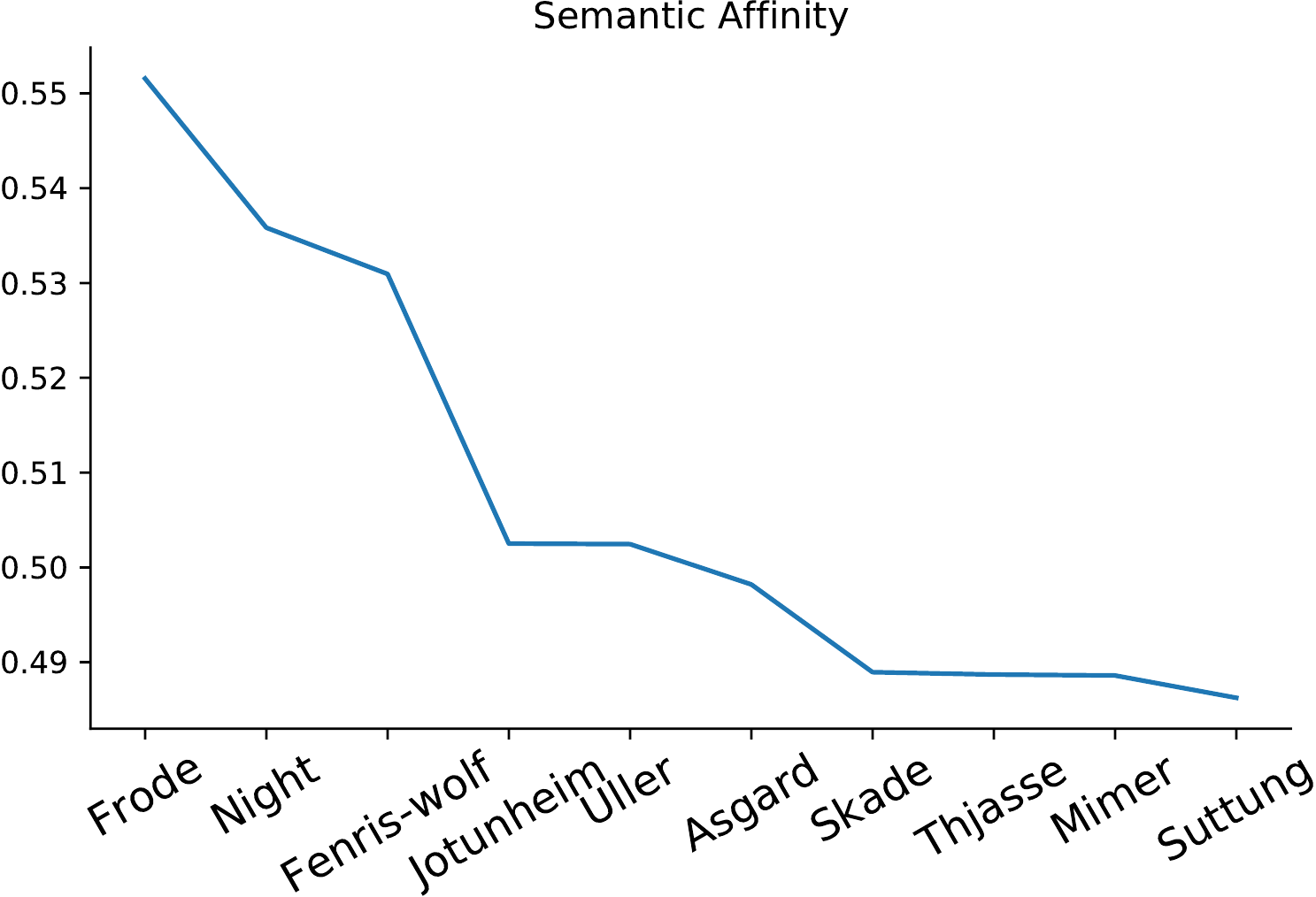} \\
		\midrule
		\multicolumn{3}{c}{Thor}   \\
		\midrule
		\includegraphics[width=0.5\linewidth]{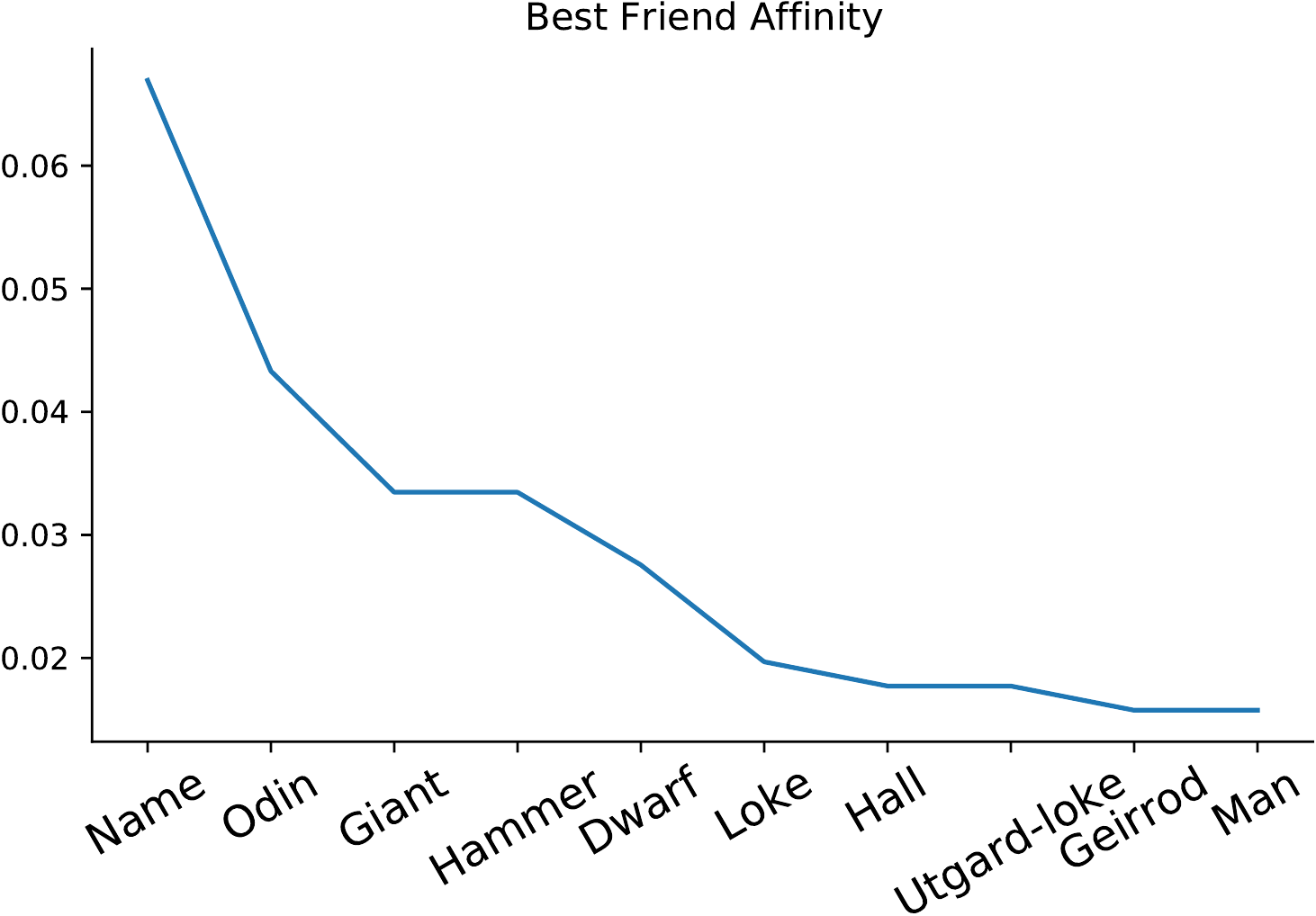} & \includegraphics[width=0.5\linewidth]{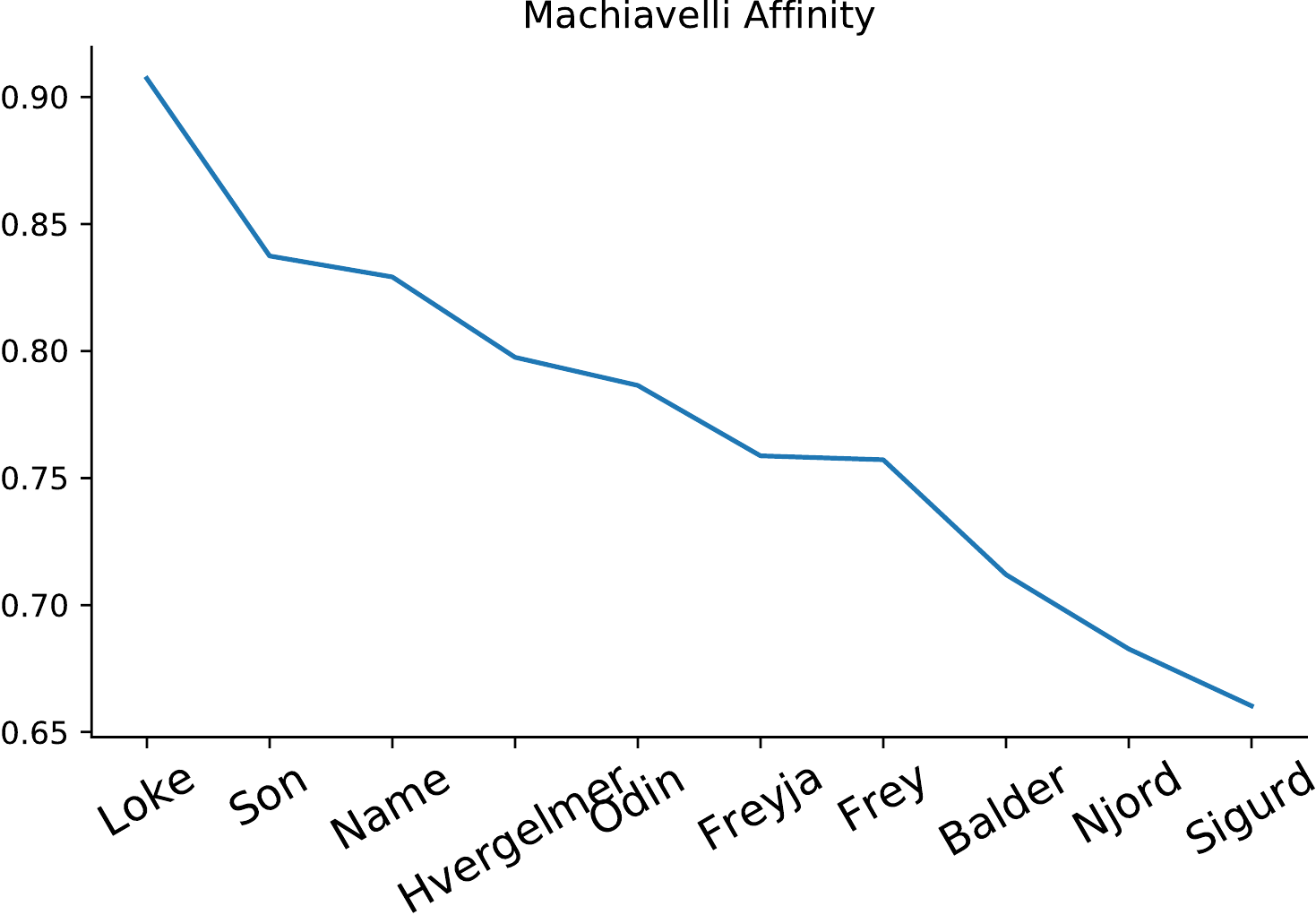} & 
		\includegraphics[width=0.5\linewidth]{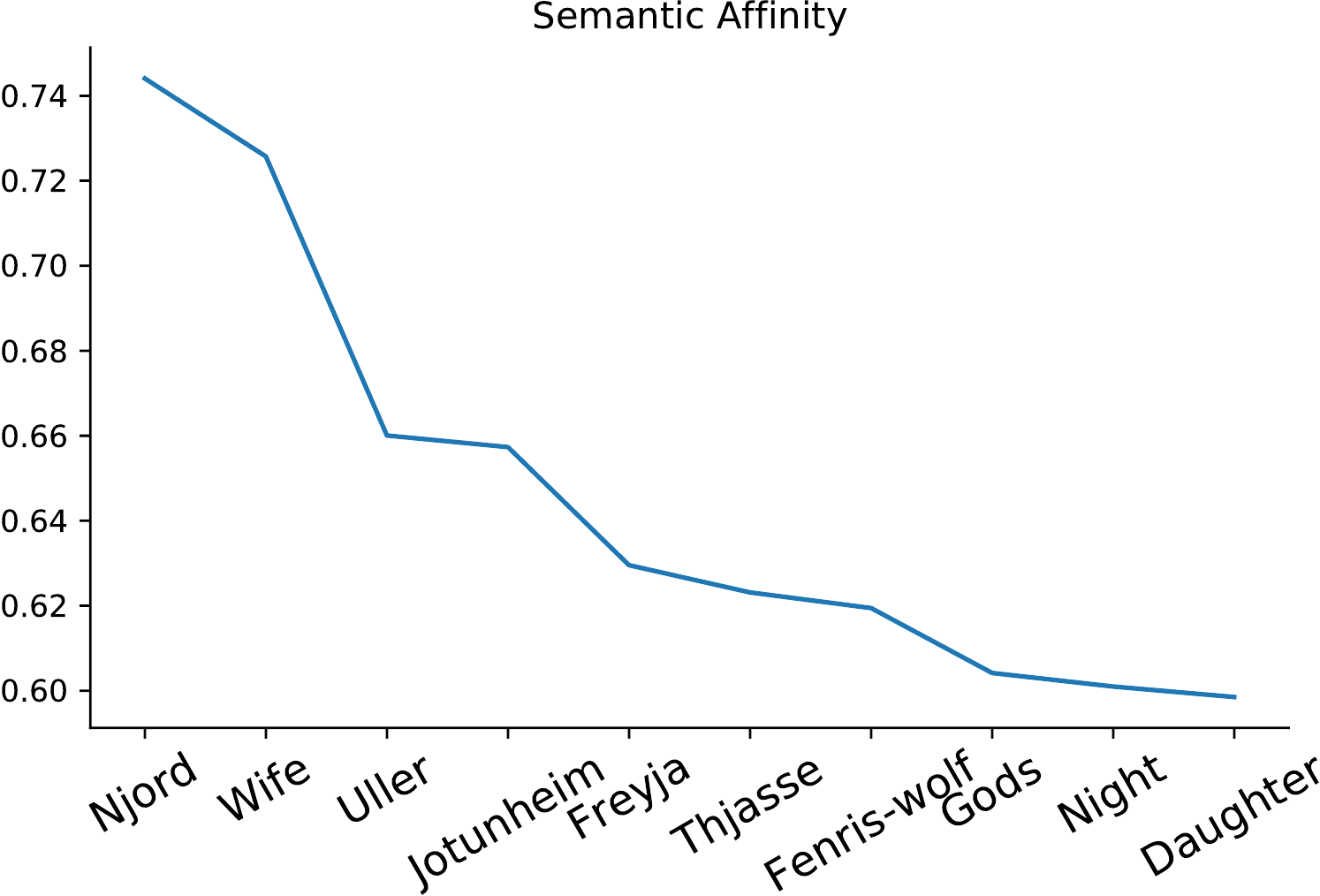} \\
		\midrule
		\multicolumn{3}{c}{Loki}   \\
		\midrule
		\includegraphics[width=0.5\linewidth]{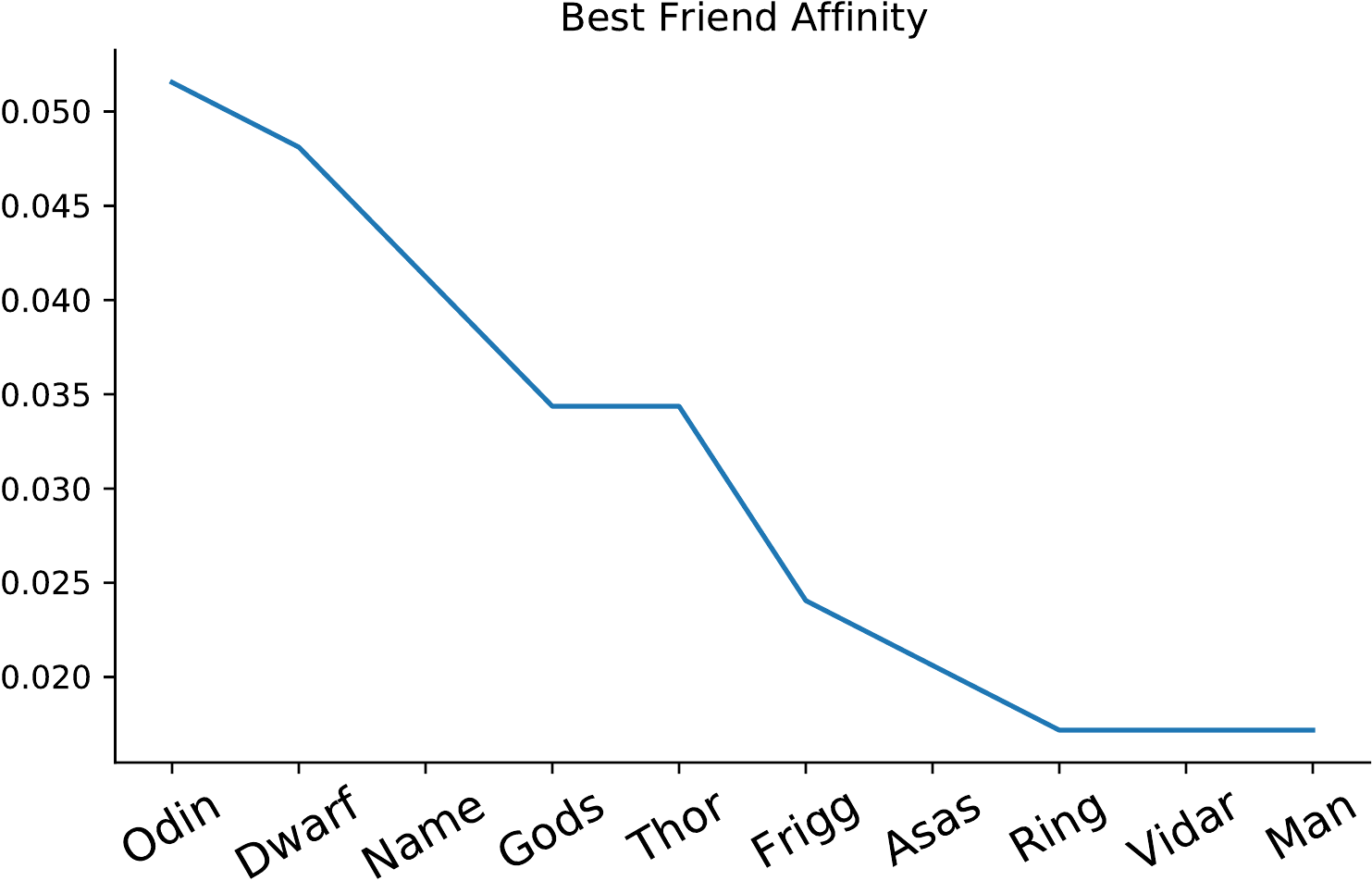} & \includegraphics[width=0.5\linewidth]{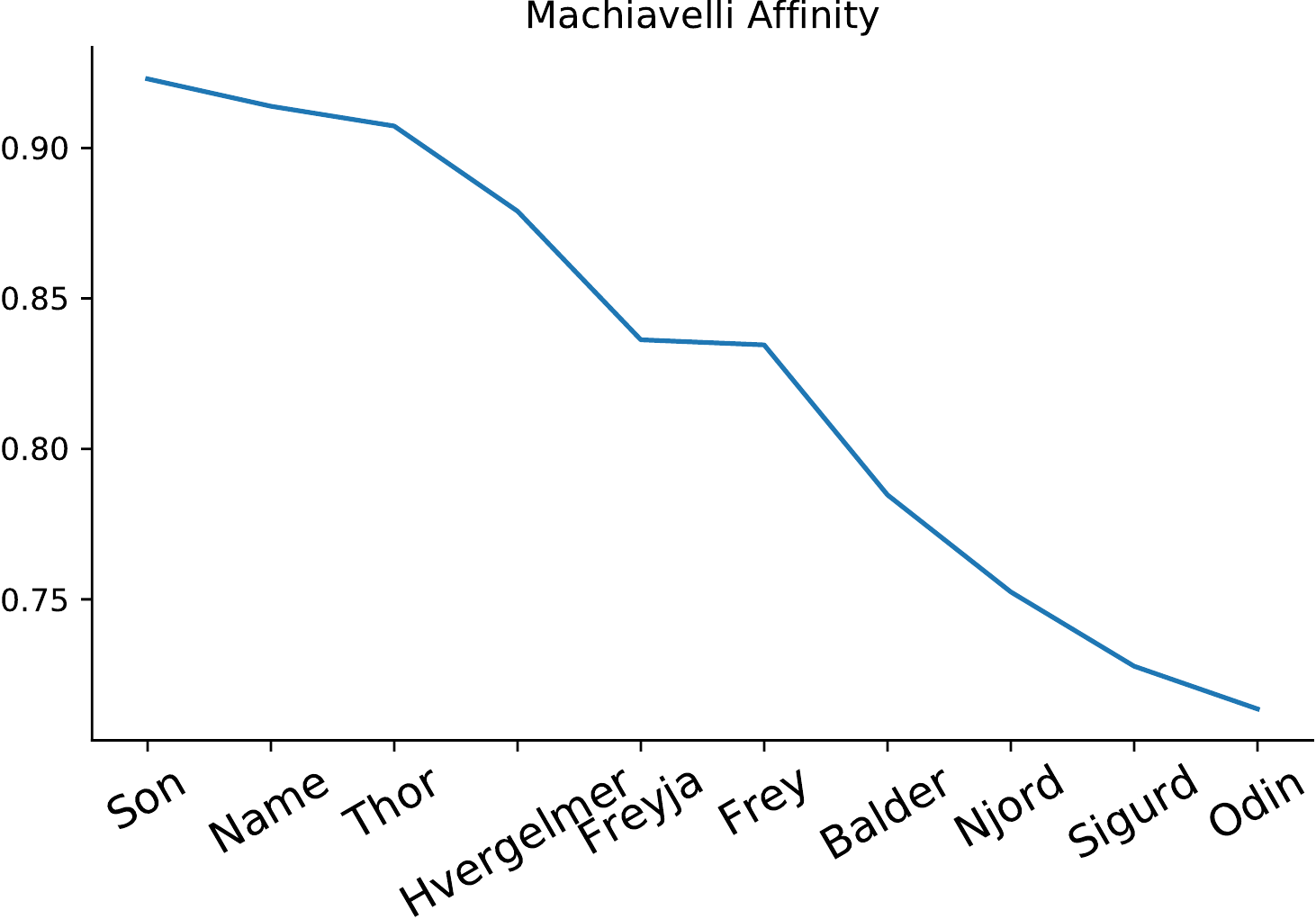} & 
		\includegraphics[width=0.5\linewidth]{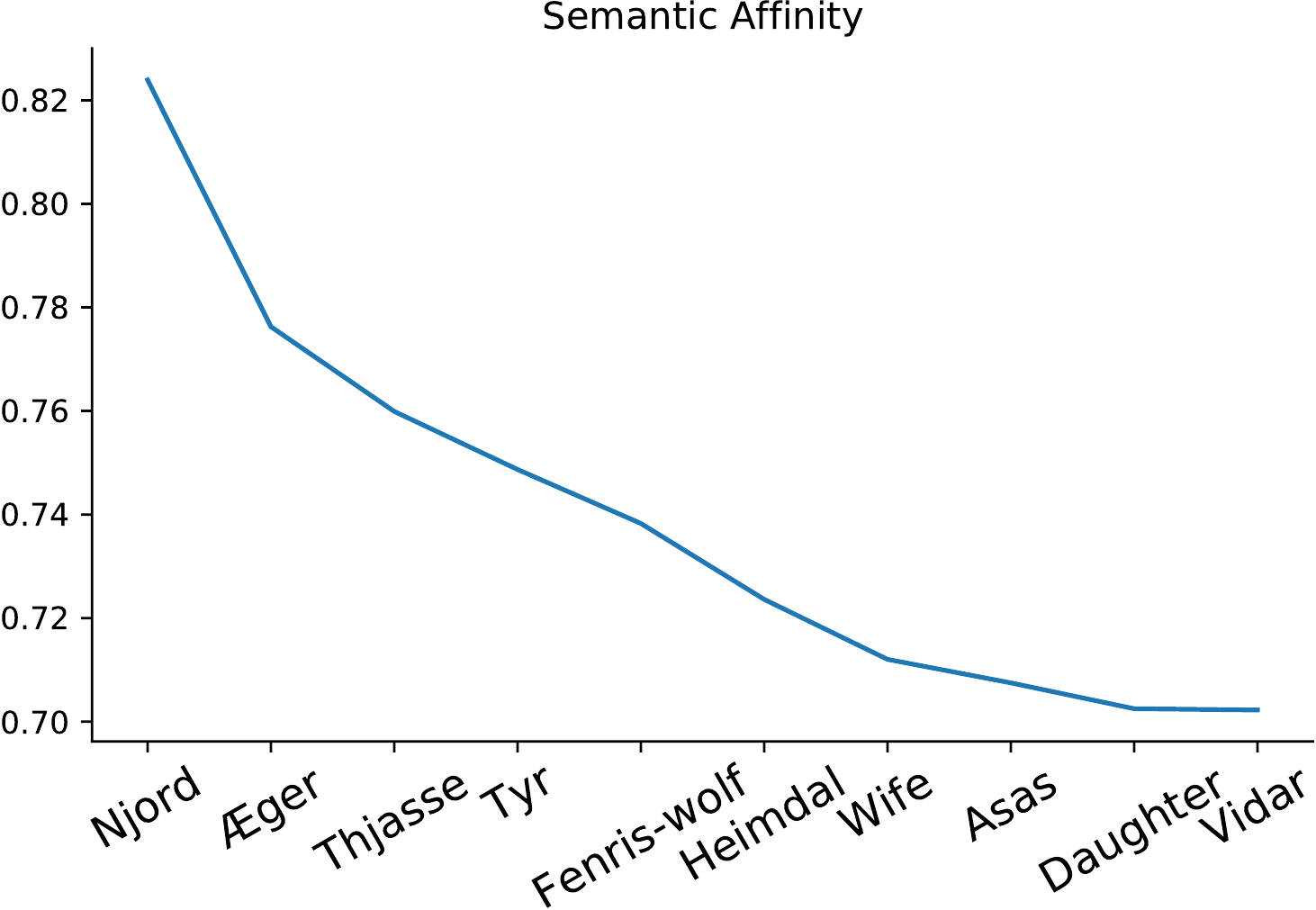} \\
		\bottomrule
	\end{tabular}
}
	\caption{\textbf{Study of affinities for three key characters in \emph{The Younger Edda}.}  Top 10 affinity values for the best friend, Machiavelli, and semantic affinity for ``Odin'', ``Thor'', and ``Loki'' in \emph{The Younger Edda} network.}
	\label{fig:odin}
\end{figure}

To complete our analysis, we have focused on the affinity study of a series of important characters in the original material. We have decided to study three characters per mythology, each one of them chosen according to their popularity and cultural significance.  

In Figure \ref{fig:zeus}  we report the results for the best friend, Machiavelli, and semantic affinities for ``Zeus'', ``Athene'', and ``Heracles''.

In the case of ``Zeus'', we can see that the best friend affinity includes mostly other Olympic gods. However, it is interesting to note that the results for the Machiavelli and semantic affinities do not show the same gods. This means that although ``Zeus'' appears repeatedly with other gods in his stories, he plays a very different role in the big picture. His highest semantic values reveal that he is mostly affine with general concepts, such as ``Sun'', ``Land'', ``Nothing'', and ``Time'', which indicates a connection between the world state and this god. The high affinities to ``Sun'' and ``King'' reinstate the connection of this god with the idea of authority.

For the case of ``Athene'', her best friend affinity mostly connects her to characters that share her stories. However, the same thing happens as with ``Zeus''. In the case of the semantic affinity, ``Phaeton'', son of the sun-god ``Helios'', is her highest affinity value, with no obvious connection between them. Following ``Midas'', ``Demeter'', is the goddess of agriculture and the afterlife, while ``Athene'' is also connected to agriculture through its association with olives and olive oil. Her top 4 semantic affinity value, ``Epimethius'', is the titan that has not foresight and he is only capable of pondering on what has already happened, which seems to be similar to wisdom, the most important attribute of ``Athene''.

``Heracles'' is the son of ``Zeus'' and ``Alcmena'', a mortal queen, and he is one of the most important heroes in the Greek mythology. Many gods appear in his top best friend affinities, as his stories are filled with Olympic deities. His top Machiavelli affinities include ``Weight'' and ``Labors'', which shows that the hero himself is tightly identified with his own narrative and his main characteristic: strength. The top 1 semantic affinity is ``Ocean''. ``Ocean'' is a titan that guarded the limitless-like waters that surrounded the known world. The connection between ``Heracles'' and the differentiation between known and uncharted waters is present in modern-day symbology because of the Pillars of Hercules, a name that the Greeks stated for the geographic formations that surrounded the strait of Gibraltar. The top 2 semantic value, the ``Hydra'', is the embodiment of chaos while the ``Heracles'' is the epitome of the Olympic order against it. This value shows us that these two characters are both impersonations of antagonistic ideas, so their connection is very strong.

We have also studied in Figure \ref{fig:danaan} the best friend, Machiavelli, and semantic affinity for three important actors in \emph{Celtic Wonder-tales}: ``Tuatha D\'{e} Danann'', ``Ireland'', and ``Lugh''.  

The ``Tuatha D\'{e} Danann'' is the pantheon of pre-Christian gods in Gaelic Ireland that populate the stories in \emph{Celtic Wonder-Tales}.  Its top best friend affinity value is ``Strength'', which reveals what is their most important attribution in these tales. The second value in best friend affinity is ``Fomor'' which is short for Fomorians. The Fomorians are the rival tribe of gods that represent the destructive face of nature, so the high affinity value reveals the importance of the conflict between the two tribes of gods in the text. 

The ``Cauldron'' is one of the four artifacts linked to the  ``Tuatha D\'{e} Danann'', alongside ``Stone'', which also appears in the top 10 of best friend values. For the Machiavelli affinity, we can see that the top value is ``Conary'', the mythical king of the Irish, seconded by ``Dagda''. In the case of the semantic affinity, the greatest value is ``Ogma''. ``Ogma'' is the god brother of ``Dagda'' and it is usually associated with him and with ``Lugh''. He is considered to be the inventor of Ogham, an ancient alphabet. This remarkable attribution, alongside the relationships with these other two gods and with his strength, makes him a very versatile god, which maybe explains why the semantic value is so high. The ``Stone'' actor is referring to the Stone of Destiny, that is the stone in which the Kings of Ireland were crowned, which connects this tribe of gods with the sovereign of the land. 

``Ireland'' is one of the most repeated concepts in these tales, in contrast to the other two mythologies, where the origin country is not so clearly stated. ``Ireland'' is being put in these tales as one of the most important topics, always associated with figures of political relevance in the mythical landscape of these tales. If we look at the best friend and Machiavelli affinities, in both cases ``Lugh'' is one of the highest affinity values. The two most important semantic affinity values are ``Stone'' and ``Danaans'', which are clearly referring to key actors in the designation of the ruler in that land as well. The third value, ``Nuada'', is the first king of the ``Tuatha D\'{e} Danann'', which reinforces even more this idea.

Finally, ``Lugh'' is one of the most important gods in Irish mythology and also a member of the ``Tuatha D\'{e} Danann''. He is the maternal grandson of ``Balor'', the leader of the Fomorians, which makes him a descendant of both tribes of gods in this mythology.  His best friend affinity values show that he is indeed tightly connected to the ``Tuatha D\'{e} Danann'' and other authoritarian symbols such as ``Ireland''. His Machiavelli affinities show that indeed the structure of actors formed around him is similar to actors that wield authority, such as ``King'' and ``Balor''. The semantic affinity reveals that the top value is ``World'', which reveals how wide the attributions and roles for this god are. ``Ildana'', ``Lauve'' and ``Fauda'' are other words to refer to ``Lugh''.


In the Nordic mythology analysis, we have opted to showcase the results for the three most popular characters of these texts: ``Odin'', ``Loki'', and ``Thor''. Their top 10 best friend, Machiavelli, and semantic affinities are collected in Figure \ref{fig:odin}.

``Odin'''s most important best friend affinities are ``Name'', as he is introduced many times in \emph{The Younger Edda} with different titles, and ``Dwarf'', as the tribes of dwarves also appear repeatedly in the presence of ``Odin'' and share many attributions.  The two most important Machiavelli affinities are his son ``Thor'' and ``Loki''. The relationship of ``Odin'' and ``Loki'' is very complex and they both take a central role in the many stories of the corpus. We also found that his highest semantic affinity is ``Frode''. ``Frode'' is another name for ``Frey'', the leader of the other tribe of gods in \emph{The Younger Edda}, the Vanir. ``Frode'' is associated with authority and sacral kingship, which explains why these gods are so affine in this case. We also found that ``Night'' has a very high semantic affinity with ``Odin''. This is indeed quite an abstract connection but it is true that ``Odin'' is highly associated with the Wild Hunt, a repeated folklorical motif in which a group of supernatural hunters lead by a mythical figure chase the skies in the night \cite{lecouteux2011phantom}. He is also considered to be a god of the dead, as he greets fallen warriors in the Nordic afterlife, and he is also capable to raise dead out of the earth. He is also affine to ``Fenris the wolf'' and ``Jotunheim'', which are enemies to ``Odin'' and key characters in the developing of the Ragnar\"{o}k.

In the case of ``Thor'', he has a high best friend affinity for ``Name'', similarly to ``Odin'', and ``Odin'' and ``Loki'' also appear in the top 10, as they are recurrent characters in his stories. For the Machiavelli affinity, ``Loki'' is again the most affine actor. In the semantic affinity, the top value is again a Vanir god, ``Njord'', god of navigation and fertility. The second value is ``Wife''. In one of \emph{The Younger Edda} tales, ``Thor'' dresses as a bride, instead of ``Freyja'', to try to recover his famous hammer, the ``Mjolnir''. This tale could be the origin of this relationship, as the semantic affinity to ``Freyja'' is also high. ``Ullr'' is the stepson of ``Thor'', associated with archery and skillfulness among other attributes, and there are evidences that he was a very important god in times before the cult of the \AE sir. There are many similarities in both of these gods attributions, which is probably the reason for this high affinity value.

For ``Loki'', the top best friend affinity is ``Odin'', followed by ``Dwarf'', because they appear repeatdly together in a set of stories of \emph{The Younger Edda}. The Machiavelli affinity values are very similar to those of ``Thor'', which emphasizes the fact that these two gods play a very similar role in the texts. The top semantic value of ``Loki'' is again ``Njord'', followed by ``\AE gir'', who is the personification of the ocean and member of the J\"{o}tun tribe, just as ``Thjasse''. ``Tyr'' is a member of the \AE sir and the god of war, whose hand is bitten by ``Fenris the wolf'', which is the next semantic value, and is also a son of ``Loki''. The presence of both J\"{o}tunn and \AE sir in the top 10 values without a clear identification with one of the tribes is a clear sign of the characteristic ambivalence of this character.

\section{Discussion and Conclusions} \label{sec:conclusions}

In this work we have studied the relationships between the meaning of different characters and concepts in three classical mythologies by extending Social Network Analysis to work with the concept of ``meaning'' or ``Noumenon'' \cite{friedman2013kant}. We have done so by combining different affinity functions and some ideas from fuzzy logic, in order to characterize the semantic value in each actor. We have also proposed a new heuristic search algorithm, the Pipe algorithm, to compute the affinity between the semantic affinity of a pair of actors, that we called the semantic affinity. This algorithm uses  a combination of affinity functions and a heuristic search in the network to compute the semantic affinity in an efficient way.

Using our proposal, we intend to model the way in which actors, words in our experimentation, enrich their own meanings by connecting with others; we can also compare the meanings of different actors or we can use them as adjectives to describe another actor. We compute a network for each culture studied, and another one obtained by fusing the three individual ones. We show the words with a more relevant meaning and the most important comparisons and similarities found, recovering already-known parallelism and similarities between them, and gaining new insights in the meaning and connotations of some of the tales studied.


  
Results in the three networks showed a mix of both historical and psychological relationships among the different actors in the network. Generally speaking, we found that gods are very close to kingship and authoritarian concepts, and particularly in the case of the Celtic myths, also to the land. We also found that gods serve as a common nexus between the different topics that appeared in each of the compilation tales, and the Nordics and Celts prefer to center their stories in gods, while the Greeks did so on humanoid and heroic figures.

When fusing all the tales together, we also observed that in the final networks three different structures, corresponding to the original tales, could be easily identified. Each one comprised mainly of the singular characters that appeared in only in of the compilations. The two most important actors that played a role of ``bridge'' between different mythologies were ``Son'' and ``Gods''. Other important actors in that sense were ``Man'', ``Father'' ``Day''. Only ``Son'' appeared in the top 10 of semantic values. Contrary to our initial hypothesis, it seems that individual characters have more semantic value than general concepts. This might be explained by the high number of attributions that many of these gods and heroes have. Logically, it is easier to tell stories about humans or human-like beings rather than essays about trades or arts, which explains why these stories fill their characters with such many different traits.

When comparing semantic affinities, we found a strong bond between kingship and the earth. This has been previously studied in many traditions thorough the world \cite{Frazer1922} and it seems that it left its footprint in these tales as well. We also found the traces of historical connections between the Nordic gods ``Loki'' and ``Odin'' and the Celtic myths. This has been hypothesized before in \cite{ewing1995birth}, and the connection is clear in this analysis. 

The main limit of this study has been the great amount of culture and relevant information that is not present in texts, but was important in the cult of each religion. This is not a problem for our new techniques, as they can work with information from outside the texts, as long as the initial network is constructed. Indeed, it is possible to construct affinity relationships using non-textual data, but it requires choosing the correct information to study and ponder carefully each source. Of course, adding more tales from different cultures is also a promising direction, but we must take into account dates, styles and original languages when doing so, or else an important deal of the findings could be more related to the translation than to the original material.

Semantic value can also be used to explore social interactions in other domains where a significant part of the information is outside the network itself, such as economic or trading networks. It also possible to use the semantic value to understand the interactions between actors in complex systems or how knowledge is propagated in networks of information \cite{bakshy2012role, chen2013information}.

\section*{Code and Data Availability}

Further results and the code to replicate our experiments can be found in this public repository: \url{https://github.com/Fuminides/noumenon_project}.

The original texts are freely available in Project Gutenberg \cite{stroube2003literary}.

\bibliographystyle{IEEEtran}
\bibliography{Semantic_Affinity}

\end{document}